%% file: main.tex
\documentclass[conference,10pt]{IEEEtran}
\IEEEoverridecommandlockouts
\usepackage{float}
\usepackage{cite}
\usepackage{amsmath,amssymb,amsfonts}
\usepackage{graphicx}
\usepackage{textcomp}
\usepackage[dvipsnames]{xcolor}
\usepackage{multirow}
\usepackage{colortbl}
\usepackage{adjustbox}
\usepackage{caption}
\usepackage{tabularx}
\usepackage{import}
\usepackage{algpseudocode,algorithm}
\usepackage{listings}
\usepackage{multirow}
\usepackage{caption}
\usepackage{subcaption}
\usepackage{hyperref}
\usepackage{circuitikz}
\usepackage{pgf}
\usepackage{pgfplots}
\usepackage{xcolor}
\usepackage{bm}
\usepackage{pgfplotstable}
\pgfplotsset{compat=1.8}
\usepackage{adjustbox}
\usepackage{changepage}

\definecolor{cardinal}{rgb}{0.6, 0.0, 0.302}
\definecolor{deeplilac}{rgb}{0.471, 0.349, 0.639}

\def\thesubsubsectiondis{\unskip\arabic{subsubsection})}
\def\BibTeX{{\rm B\kern-.05em{\sc i\kern-.025em b}\kern-.08em
    T\kern-.1667em\lower.7ex\hbox{E}\kern-.125emX}}

\newcolumntype{Y}{>{\centering\arraybackslash}X}
    
\usepackage{paralist}

\usepackage{listings}

\begin{document}
\include{macros}
\newcolumntype{Y}{>{\centering\arraybackslash}X}

\title{\huge
% Trojan Resistant Design-for-Trust via Rareness Reduction\\
% Towards Design-for-Trust using Rareness Reduction \\
% Rareness Reduction for Design-for-Trust \\
Design for Trust Utilizing Rareness Reduction\\
% Rareness Reduction for Easier Trojan Detection \\
% \vspace{-12mm}
\thanks{This work was partially supported by the NSF grant CCF-1908131.}
}
\author{Aruna Jayasena and Prabhat Mishra\\
Department of Computer \& Information Science \& Engineering\\
University of Florida, Gainesville, Florida, USA}

\iffalse
\author{\IEEEauthorblockN{1\textsuperscript{st} Given Name Surname}
\IEEEauthorblockA{\textit{dept. name of organization (of Aff.)} \\
\textit{name of organization (of Aff.)}\\
City, Country \\
email address or ORCID}
\and
\IEEEauthorblockN{2\textsuperscript{nd} Given Name Surname}
\IEEEauthorblockA{\textit{dept. name of organization (of Aff.)} \\
\textit{name of organization (of Aff.)}\\
City, Country \\
email address or ORCID}
\and
\IEEEauthorblockN{3\textsuperscript{rd} Given Name Surname}
\IEEEauthorblockA{\textit{dept. name of organization (of Aff.)} \\
\textit{name of organization (of Aff.)}\\
City, Country \\
email address or ORCID}
\and
\IEEEauthorblockN{4\textsuperscript{th} Given Name Surname}
\IEEEauthorblockA{\textit{dept. name of organization (of Aff.)} \\
\textit{name of organization (of Aff.)}\\
City, Country \\
email address or ORCID}
\and
\IEEEauthorblockN{5\textsuperscript{th} Given Name Surname}
\IEEEauthorblockA{\textit{dept. name of organization (of Aff.)} \\
\textit{name of organization (of Aff.)}\\
City, Country \\
email address or ORCID}
\and
\IEEEauthorblockN{6\textsuperscript{th} Given Name Surname}
\IEEEauthorblockA{\textit{dept. name of organization (of Aff.)} \\
\textit{name of organization (of Aff.)}\\
City, Country \\
email address or ORCID}
}
\fi

\maketitle
\vspace{-1in}
\begin{abstract}
Increasing design complexity and reduced time-to-market have motivated manufacturers to outsource some parts of the System-on-Chip (SoC) design flow to third-party vendors. This provides an opportunity for attackers to introduce hardware Trojans by constructing stealthy triggers consisting of rare events (e.g., rare signals, states, and transitions). There are promising test generation-based hardware Trojan detection techniques that rely on the activation of rare events. In this paper, we investigate rareness reduction as a design-for-trust solution to make it harder for an adversary to hide Trojans (easier for Trojan detection). Specifically, we analyze different avenues to reduce the potential rare trigger cases, including design diversity and area optimization.
While there is a good understanding of the relationship between area, power, energy, and performance, this research provides a better insight into the dependency between area and security. Our experimental evaluation demonstrates that area reduction leads to a reduction in rareness. It also reveals that reducing rareness leads to faster Trojan detection as well as improved coverage by Trojan detection methods. 
\end{abstract}


\ctikzset{
    logic ports=ieee,
    logic ports/scale=0.6,
    % logic ports/fill=lightgray
}

\import{sections/}{introduction.tex}

\import{sections/}{related.tex}
\import{sections/}{methodology.tex}

\import{sections/}{experiments.tex}

\import{sections/}{conclusion.tex}

\bibliographystyle{IEEEtran}
\bibliography{IEEEabrv,bibliography}

\end{document}

%% file: macros.tex
% =========================================================================

% General

\newlength{\savejot}
\setlength{\savejot}{\jot}

\newcommand{\xor}{{\,\oplus\,}}
\newcommand{\bits}{{\{0,1\}}}

\newcommand{\heading}[1]{{\vspace{10pt}\noindent{\textsc{#1}}}}
\newcommand{\noskipheading}[1]{{\noindent{\textsc{#1}}}}

\newcommand{\headingg}[1]{{\textsc{#1}}}
\newcommand{\Heading}[1]{{\vspace{8pt}\noindent\textbf{#1}}}

\newcommand{\emptystring}{\varepsilon}
\newcommand{\concat}{\:\|\:}
\newcommand{\Concat}{\;\|\;}
\newcommand{\Colon}{{\,:\;\,}}

\newcommand{\N}{{{\sf N}}}
\newcommand{\R}{{{\rm\bf R}}}
\newcommand{\Y}{{{\sf Y}}}
\newcommand{\Z}{{{Z}}}

\newcommand{\calC}{{\cal C}}
\newcommand{\calD}{{\cal D}}
\newcommand{\calE}{{\cal E}}
\newcommand{\calF}{{\cal F}}
\newcommand{\calI}{{\cal I}}
\newcommand{\calO}{{\cal O}}
\newcommand{\calQ}{{\cal Q}}
\newcommand{\calR}{{\cal R}}
\newcommand{\calG}{{\cal G}}
\newcommand{\calK}{{\cal K}}
\newcommand{\calN}{{\cal N}}
\newcommand{\calT}{{\cal T}}
\newcommand{\calH}{{\mathcal{H}}}
\newcommand{\calM}{{\mathcal{M}}}
\newcommand{\calV}{{\mathcal{V}}}
\newcommand{\calP}{{\mathcal{P}}}

\newcommand{\D}{{\calD}}
\newcommand{\E}{{\calE}}
\newcommand{\K}{{\calK}}
\newcommand{\I}{{\calI}}
\renewcommand{\R}{{\calR}}
\renewcommand{\O}{{\calO}}

\newcommand{\goesto}{{\rightarrow}}
\newcommand{\eqdef}{\stackrel{\rm def}{=}}
\newcommand{\getsr}{{\:\stackrel{{\scriptscriptstyle \hspace{0.2em}\$}} {\leftarrow}\:}}
\newcommand{\getsd}{{\:\stackrel{{\scriptscriptstyle \hspace{0.2em}D}} {\leftarrow}\:}}
\renewcommand{\choose}[2]{{{#1}\atopwithdelims(){#2}}}
\newcommand{\abs}[1]{{\displaystyle \left| {#1} \right| }}
\newcommand{\EE}[1]{{\E\left[{#1}\right]}}

\newcommand{\Damgard}{{\mbox{Damg{\aa}rd}}}

\newcommand{\strtonum}[1]{{ \mathsf{str2num}\left({#1}\right) }}
\newcommand{\numtostr}[2]{{ \mathsf{num2str}_{\scriptscriptstyle #2}
\left( {#1} \right) }}

\newcommand{\Adv}{{\mathbf{\bf Adv}}}

\newcommand{\Randword}{\mathrm{Rand}}
\newcommand{\Permword}{\mathrm{Perm}}
\newcommand{\Randd}[2]{\ensuremath{\Randword({#1},{#2})}}
\newcommand{\Rand}[1]{\ensuremath{\Randword({#1})}}
\newcommand{\Rn}{{\calR_n}}
\newcommand{\Pn}{{\calP_n}}
\newcommand{\Perm}[1]{\ensuremath{\Permword({#1})}}
\newcommand{\Permm}[2]{\ensuremath{\Permword({#1},{#2})}}

\newcommand{\GF}{{\mathrm{GF}}}

\newcommand{\code}[1]{{\langle{#1}\rangle}}

\newcommand{\then}{{;\;}}
\newcommand{\andthen}{{\::\;\;}}

\newcommand{\ie}{i.e.}
\newcommand{\eg}{e.g.}
\newcommand{\adv}{\ensuremath{\mathrm{Adv}}}
\newcommand{\hdir}{\mbox{\vspace{6mm}\~{ }}}

% ========================================================================

%% file: sections/introduction.tex
\section{Introduction} \label{sec:introduction}
The complexity of the hardware designs continues to grow over the years. To make matters worse, the hardware development life cycle has been shortened significantly. As a consequence, the designers do not have enough time to verify the functional behaviors as well as non-functional (e.g., security) requirements. This opens up opportunities for attackers to implant malicious circuits into the designs that can lead to serious security risks. This research utilizes rareness reduction techniques to improve the security verification process to enable trustworthy hardware systems.

\subsection{Threat Model}\label{sec:threat}
\begin{figure}[htp]
    \begin{center}
    \begin{adjustwidth*}{1em}{}
        \input{Images/trojan}
    \end{adjustwidth*}
    \end{center}
     \vspace{-0.1in}
      \caption{Hardware Trojan triggered by two rare signals (p,q)}
      \vspace{-0.15in}
  \label{fig:Trojan}
\end{figure}
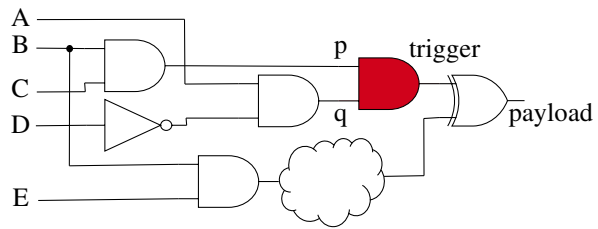

We consider the threat model under supply chain vulnerability where the attackers (untrusted foundry, rogue designer, malicious CAD tool \cite{polian2016trojans}) can insert stealthy hardware Trojans that can stay hidden during traditional functional validation and testing. Specifically, an attacker is likely to combine several rare signals with low activation probabilities as the trigger for the Trojan. Once the Trojan is activated, it may alter the functionality, leak sensitive data to the outputs, or perform other malicious activities. Figure~\ref{fig:Trojan} shows a simple hardware Trojan that is triggered by two rare signals of $p$ and $q$, while it flips the design output as the payload.
% Our threat model assumes that an adversary can introduce hardware Trojans (HT) during any stages of the SoC development cycle including the design of RTL models, synthesis to gate-level netlist, and fabrication. An HT consists of two major components: trigger and payload. An adversary is likely to construct the trigger such that it will avoid detection during traditional validation using millions of random or constrained-random test patterns. A stealthy trigger can be constructed using rare events such as rare signals (states) or rare branches (transitions). Figure~\ref{fig:Trojan} shows an example Trojan circuit with a trigger and payload. When the trigger gets activated, the payload can enable malicious activities such as information leakage, incorrect execution, or denial-of-service.  
% In the example circuit, the Trojan payload flips the expected output.

% \vspace{-0.1in}

\subsection{Limitations of Existing Methods}\label{sec:limits}
% There are logic encryption techniques proposed in the literature for reducing the rareness of signals in hardware designs~\cite{samimi2016hardware, dupuis2014novel,marcelli2017evolutionary}.

Hardware obfuscation (logic encryption) is a promising avenue to build a design-for-trust solution since it is hard for the attacker to figure out the functionality without the key, and therefore, hard to identify where to hide the Trojans~\cite{samimi2016hardware, dupuis2014novel,marcelli2017evolutionary}. However, it can lead to unacceptable area, power and performance overhead. Samimi et al. used logic obfuscation to reduce rareness of signals~\cite{samimi2016hardware}. However, it faces three practical limitations. First, it inherits the disadvantages of obfuscation  and leads to significant area (32\%), power (9\%), and performance (56\%) overhead. Next, the requirement of a key itself can be under attack. Finally, the applicability is limited to small combinational designs (applied on simple designs with less than 4000 gates). 

There are promising research efforts for efficient detection of hardware Trojans that can be broadly divided into the following categories: statistical test generation methods~\cite{mero2009swarup,Mers2016yuanwen}, directed test generation methods~\cite{tarmac2021yangdi}, machine learning based techniques, side-channel analysis based techniques~\cite{fournaris2018fpga}, self-referencing based techniques, and also equivalence checking based techniques~\cite{bhasin2015survey}. The success of these methods heavily depend on the number of potential triggers for Trojans in the hardware designs. Specifically, if there are too many rare signals or a lot of rare signals with very low probabilities, it would be infeasible for the existing methods to detect any stealthy Trojans constructed from these rare signals. The proposed rareness reduction will be helpful for existing Trojan detection methods, as demonstrated in Section~\ref{subsec:exp:testgeneration}. 

\subsection{Research Contributions}\label{sec:contribution}
In this paper, we look at Trojan detection problem from an orthogonal perspective. We try to eliminate the hiding places for Trojans as much as possible during the design stage. This contributes to design-for-trust from two complementary avenues. (1) The reduced rareness can demotivate the attackers to introduce malicious implants in the design due to less number of potential triggers. (2) Trojan detection approaches can take the benefit of reduced rareness for faster and efficient Trojan detection. Specifically, this papers makes the following major contributions.
\begin{itemize}
    \item We perform a theoretical analysis of the root causes of rare signals that are likely to be exploited by adversaries to construct stealthy triggers in hardware Trojans.
    \item We explore various methods for rareness reduction, including design diversity and area optimization.
%    \item Although there is a clear understanding of the area, power, and energy ($A\uparrow$ $>$ $P\uparrow$) of hardware designs and how it impacts the execution time because of the relationship of $P = E/t$. However, there is no such clear understanding of the area and security of hardware designs.
    \item To the best of our knowledge, this is the first attempt in formulating a theoretical relationship between design area and hardware security, and confirming with empirical results on real-world hardware designs.
    \item Experimental evaluation demonstrates the effectiveness of rareness reduction for Trojan detection using statistical test generation as well as maximal clique activation.  
\end{itemize}
  
%Minimizing the rareness can affect the security of the design in several ways. (i) It reduces the security verification time due to the reduction of potential trigger points, (ii) validation effort gets reduced due to the increment of the average signal rareness, and (ii) improves the coverage of the verification techniques.

% \begin{enumerate}
%     \item \textit{{What is the rationale between rare signals in a design?}} 
%       \vspace{0.05in}
       
%     \noindent Rareness of a signal depends on how deep the signal is in the design. 
%     \vspace{0.05in}

%     \item \textit{{Does design area affects the rareness of the signals in the design?}} 
%       \vspace{0.05in}
       
%     \noindent 
%         \vspace{0.05in}
        
%     \item \textit{{Can we reduce the rareness of signals in a design?}} 
%       \vspace{0.05in}
       
%     \noindent 
%         \vspace{0.05in}
        
%     \item \textit{{How rareness reduction affects the test generation and Trojan detection process?}} 
%       \vspace{0.05in}
       
%     \noindent 
%         \vspace{0.05in}

%  \end{enumerate}
 
This paper is organized as follows. Section~\ref{sec:methodology} describes our proposed methodology.
Section~\ref{sec:experiments} presents the experimental results. Section~\ref{sec:conclusion} concludes the paper.

%% file: Images/trojan.tex
\begin{tikzpicture}[x=0.75pt,y=0.75pt,yscale=-.9,xscale=0.8]
    %Shape: And Gate [id:dp5940246098916848] 
    \draw   (109.71,25.08) -- (128.7,25.08) .. controls (139.18,25.08) and (147.69,31.57) .. (147.69,39.55) .. controls (147.69,47.54) and (139.18,54.03) .. (128.7,54.03) -- (109.71,54.03) -- (109.71,25.08) -- cycle (97.06,29.91) -- (109.71,29.91) (97.06,49.2) -- (109.71,49.2) (147.69,39.55) -- (160.35,39.55) ;
    %Shape: And Gate [id:dp4902057915113851] 
    \draw   (168.78,90.02) -- (187.77,90.02) .. controls (198.25,90.02) and (206.76,96.5) .. (206.76,104.49) .. controls (206.76,112.47) and (198.25,118.96) .. (187.77,118.96) -- (168.78,118.96) -- (168.78,90.02) -- cycle (156.12,94.84) -- (168.78,94.84) (156.12,114.13) -- (168.78,114.13) (206.76,104.49) -- (219.42,104.49) ;
    %Shape: And Gate [id:dp10735905081000441] 
    \draw  [fill={rgb, 255:red, 255; green, 255; blue, 255 }  ,fill opacity=1 ] (206.62,44.8) -- (225.61,44.8) .. controls (236.09,44.8) and (244.59,51.28) .. (244.59,59.27) .. controls (244.59,67.26) and (236.09,73.74) .. (225.61,73.74) -- (206.62,73.74) -- (206.62,44.8) -- cycle (193.96,49.62) -- (206.62,49.62) (193.96,68.92) -- (206.62,68.92) (244.59,59.27) -- (257.25,59.27) ;
    %Straight Lines [id:da6014034747793278] 
    \draw    (97.06,49.2) -- (97.46,54.1) ;
    %Straight Lines [id:da7230459139740377] 
    \draw    (160.35,39.55) -- (257.25,39.97) ;
    %Straight Lines [id:da30186153579757713] 
    \draw    (66.63,29.67) -- (97.06,29.91) ;
    %Straight Lines [id:da4039028189041829] 
    \draw    (311,101.2) -- (285.8,101.6) ;
    %Straight Lines [id:da7977605893356641] 
    \draw    (161,68.91) -- (193.96,68.92) ;
    %Shape: Ellipse [id:dp4118443439707664] 
    \draw  [fill={rgb, 255:red, 0; green, 0; blue, 0 }  ,fill opacity=1 ][line width=0.75]  (86.1,29.97) .. controls (86.1,29.14) and (86.77,28.47) .. (87.6,28.47) .. controls (88.43,28.47) and (89.1,29.14) .. (89.1,29.97) .. controls (89.1,30.8) and (88.43,31.47) .. (87.6,31.47) .. controls (86.77,31.47) and (86.1,30.8) .. (86.1,29.97) -- cycle ;
    %Straight Lines [id:da45384623138520386] 
    \draw    (310.87,69.14) -- (320.55,68.61) ;
    %Straight Lines [id:da705753086763653] 
    \draw    (310.87,69.14) -- (311,101.2) ;
    %Straight Lines [id:da08859978158834614] 
    \draw    (66.31,73.22) -- (98.08,72.93) ;
    %Straight Lines [id:da826280050687314] 
    \draw    (66.88,54.27) -- (97.46,54.1) ;
    %Shape: Not/Inverter Gate [id:dp3725847724285447] 
    \draw   (109.8,58.46) -- (144.96,72.93) -- (109.8,87.4) -- (109.8,58.46) -- cycle (98.08,72.93) -- (109.8,72.93) (151.99,72.93) -- (161.37,72.93) (144.96,72.93) .. controls (144.96,71.33) and (146.53,70.03) .. (148.48,70.03) .. controls (150.42,70.03) and (151.99,71.33) .. (151.99,72.93) .. controls (151.99,74.53) and (150.42,75.82) .. (148.48,75.82) .. controls (146.53,75.82) and (144.96,74.53) .. (144.96,72.93) -- cycle ;
    %Straight Lines [id:da9584543079757086] 
    \draw    (67.78,115.11) -- (156.12,114.13) ;
    %Straight Lines [id:da6510075389320418] 
    \draw    (87.4,95.67) -- (156.12,94.84) ;
    %Straight Lines [id:da8703792752424429] 
    \draw    (87.4,95.67) -- (87.6,30.47) ;
    %Straight Lines [id:da3118285199934787] 
    \draw    (160.27,49.82) -- (160,13.5) ;
    %Straight Lines [id:da8229782896454436] 
    \draw    (193.96,49.62) -- (160.27,49.82) ;
    %Shape: And Gate [id:dp3021120518500413] 
    \draw  [fill={rgb, 255:red, 208; green, 2; blue, 27 }  ,fill opacity=0.58 ] (269.91,35.15) -- (288.9,35.15) .. controls (299.38,35.15) and (307.89,41.63) .. (307.89,49.62) .. controls (307.89,57.61) and (299.38,64.09) .. (288.9,64.09) -- (269.91,64.09) -- (269.91,35.15) -- cycle (257.25,39.97) -- (269.91,39.97) (257.25,59.27) -- (269.91,59.27) (307.89,49.62) -- (320.55,49.62) ;
    %Shape: Xor Gate [id:dp6829762489291851] 
    \draw   (328.62,44.87) -- (342.08,44.87) .. controls (351.47,45.13) and (359.86,50.68) .. (363.61,59.12) .. controls (359.86,67.55) and (351.47,73.11) .. (342.08,73.36) -- (328.62,73.36) .. controls (334.39,64.55) and (334.39,53.69) .. (328.62,44.87) -- cycle (320.55,49.62) -- (331.31,49.62) (320.55,68.61) -- (331.31,68.61) (363.61,59.12) -- (374.38,59.12) (325.93,44.87) .. controls (331.7,53.69) and (331.7,64.55) .. (325.93,73.36) ;
    %Shape: Cloud [id:dp30006075769215257] 
    \draw   (225.82,96.66) .. controls (225.29,92.68) and (227.04,88.74) .. (230.33,86.51) .. controls (233.62,84.28) and (237.88,84.15) .. (241.29,86.19) .. controls (242.5,83.87) and (244.71,82.27) .. (247.26,81.87) .. controls (249.8,81.47) and (252.39,82.32) .. (254.22,84.16) .. controls (255.25,82.06) and (257.27,80.65) .. (259.57,80.43) .. controls (261.87,80.21) and (264.11,81.21) .. (265.51,83.08) .. controls (267.37,80.85) and (270.33,79.91) .. (273.11,80.67) .. controls (275.89,81.43) and (277.99,83.74) .. (278.5,86.61) .. controls (280.78,87.25) and (282.68,88.85) .. (283.7,91.02) .. controls (284.73,93.19) and (284.78,95.7) .. (283.85,97.91) .. controls (286.1,100.88) and (286.62,104.84) .. (285.23,108.31) .. controls (283.84,111.77) and (280.75,114.23) .. (277.1,114.76) .. controls (277.07,118.02) and (275.32,121) .. (272.51,122.57) .. controls (269.7,124.14) and (266.28,124.04) .. (263.56,122.32) .. controls (262.4,126.22) and (259.14,129.09) .. (255.19,129.69) .. controls (251.23,130.28) and (247.3,128.51) .. (245.08,125.12) .. controls (242.35,126.79) and (239.09,127.27) .. (236.01,126.45) .. controls (232.94,125.64) and (230.32,123.59) .. (228.74,120.78) .. controls (225.96,121.11) and (223.27,119.65) .. (222.01,117.11) .. controls (220.74,114.58) and (221.18,111.52) .. (223.09,109.44) .. controls (220.61,107.96) and (219.34,105.01) .. (219.95,102.14) .. controls (220.56,99.27) and (222.91,97.12) .. (225.77,96.82) ; \draw   (223.09,109.44) .. controls (224.26,110.14) and (225.61,110.46) .. (226.97,110.35)(228.74,120.78) .. controls (229.32,120.71) and (229.89,120.57) .. (230.44,120.35)(245.08,125.12) .. controls (244.67,124.5) and (244.32,123.83) .. (244.05,123.13)(263.56,122.32) .. controls (263.77,121.61) and (263.91,120.88) .. (263.97,120.14)(277.1,114.76) .. controls (277.13,111.3) and (275.19,108.12) .. (272.12,106.61)(283.85,97.91) .. controls (283.36,99.09) and (282.6,100.14) .. (281.64,100.97)(278.5,86.61) .. controls (278.58,87.09) and (278.62,87.57) .. (278.61,88.06)(265.51,83.08) .. controls (265.05,83.63) and (264.67,84.25) .. (264.38,84.92)(254.22,84.16) .. controls (253.98,84.67) and (253.79,85.2) .. (253.67,85.75)(241.29,86.19) .. controls (242.01,86.62) and (242.68,87.13) .. (243.28,87.73)(225.82,96.66) .. controls (225.9,97.21) and (226.01,97.76) .. (226.17,98.29) ;
    %Straight Lines [id:da7377809732165184] 
    \draw    (66.96,13.67) -- (160,13.5) ;
    %Straight Lines [id:da38540160613140384] 
    \draw    (161.37,72.93) -- (161,68.91) ;
    
    % Text Node
    \draw (61.26,12.12) node   [align=left] {\begin{minipage}[lt]{12.68pt}\setlength\topsep{0pt}
    A
    \end{minipage}};
    % Text Node
    \draw (61.52,51.93) node   [align=left] {\begin{minipage}[lt]{12.68pt}\setlength\topsep{0pt}
    C
    \end{minipage}};
    % Text Node
    \draw (61.78,27.5) node   [align=left] {\begin{minipage}[lt]{12.68pt}\setlength\topsep{0pt}
    B
    \end{minipage}};
    % Text Node
    \draw (61.26,72.15) node   [align=left] {\begin{minipage}[lt]{12.68pt}\setlength\topsep{0pt}
    D
    \end{minipage}};
    % Text Node
    \draw (62.26,113.15) node   [align=left] {\begin{minipage}[lt]{12.68pt}\setlength\topsep{0pt}
    E
    \end{minipage}};
    % Text Node
    \draw (265,30.78) node   [align=left] {\begin{minipage}[lt]{12.68pt}\setlength\topsep{0pt}
    p
    \end{minipage}};
    % Text Node
    \draw (265,69) node   [align=left] {\begin{minipage}[lt]{12.68pt}\setlength\topsep{0pt}
    q
    \end{minipage}};
    % Text Node
    \draw (312.26,31) node   [align=left] {\begin{minipage}[lt]{12.68pt}\setlength\topsep{0pt}
    trigger
    \end{minipage}};
    % Text Node
    \draw (375,67) node   [align=left] {\begin{minipage}[lt]{12.68pt}\setlength\topsep{0pt}
    payload
    \end{minipage}};

    \end{tikzpicture}

%% file: sections/methodology.tex
\section{Rareness Reduction}\label{sec:methodology}
%\vspace{-0.1in}

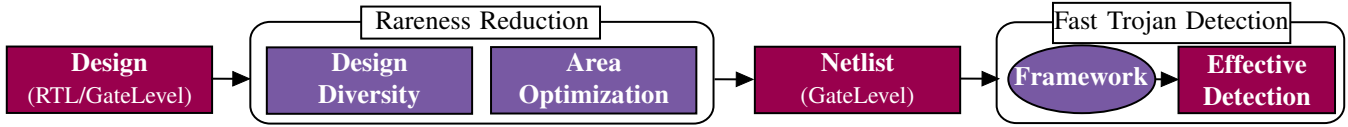
\begin{figure*}[htp]
        % \vspace{-0.1in}
    \begin{center}
    \input{Images/overview}    
    \end{center}

      \vspace{-0.1in}
      \caption{Overview of proposed rareness reduction based design-for-trust improvement.}
        \vspace{-0.2in}
      \label{fig:overview}
\end{figure*}

In this section, we perform theoretical analysis as well as exploration of rareness reduction techniques. Specifically, this section is organized as follows. First, we define few terms that are used in the rest of the paper. Next, we introduce metrics to compare rareness between designs. Then, we perform a  theoretical analysis of the root causes of the rare signals in hardware designs. We also explore several rareness reduction techniques. Finally, we discuss the effect of rareness reduction on two state-of-the-art Trojan detection techniques.

\subsection{Definitions}
We define three terms that are used in the rest of the paper.

\vspace{0.05in}
\noindent \textbf{Definition 1:} \textit{Rareness of a Signal} ($S_\omega$)

\noindent Every signal has two possible values: high (`1') and low (`0'). 
We define the rareness of a signal $S$ as the minimum of the two probabilities as shown below. For example, if signal $S^i$ is `0' 10\% of the time (`1' for 90\% of the time) during simulation, $S_\omega^i$ is 0.1. 
\vspace{-0.1in}
\begin{equation}\footnotesize
    S_\omega = \min({P(S\gets0),P(S\gets1)})
\end{equation}
 \vspace{-0.2in}

\noindent \textbf{Definition 2:} \textit{Logic Probability Vector} ($P(\bar S)$)

\noindent In order to represent the probabilities of a signal $S$ having a value ``Low" (``0") and ``High" (``1"), we use the following vector and matrix representations.
\vspace{-0.1in}
\begin{equation}\footnotesize
    P(\bar S) = <P(0),P(1)> =\begin{bmatrix}
P(0) & 0 \\
0 & P(1) 
\end{bmatrix}
\end{equation}
\vspace{-0.2in}

\noindent \textbf{Definition 3:} \textit{Ideal Transfer Matrix}

\noindent The Ideal Transfer Matrix (ITM) is used for the reliability evaluation of logic circuits~\cite{franco2008signal}. In ITM, we express the truth table of a logic gate in matrix representation where rows represent the inputs combinations of the gate while two columns represent the output signal being the value of 0 and 1. ITM representation of primary logic gate types (AND,OR,NOT) is shown in Equation~\ref{eqn:itmgates}. ITM representation for other gates can be computed in a similar way. 
%  In such a matrix M for an n-input, m-output logic function f , each entry M(i, j),
% with binary indices i = i0i1 ...in−1, j = j0 j1 ... jm−1, contains a 1 if the binary
% inputs i0, i1, ...in−1 produce the binary output values j0, j1, ... jm−1, and contains a 0 otherwise. This representation is referred to here as an ideal transfer
% matrix (ITM).

\begin{figure}[H]
    \centering
    \footnotesize
    \vspace{-0.3in}
    \begin{equation}
        ITM_{AND} =  \begin{bmatrix}
        1 & 0 \\
        1 & 0 \\
        1 & 0 \\
        0 & 1
        \end{bmatrix} 
        ITM_{OR} =  \begin{bmatrix}
        0 & 1 \\
        0 & 1 \\
        0 & 1 \\
        1 & 0
        \end{bmatrix} 
        ITM_{NOT} =  \begin{bmatrix}
        0 & 1 \\
        1 & 0 \\
        \end{bmatrix} 
        \label{eqn:itmgates}
   \end{equation}
  \vspace{-0.2in}
    % \caption{Caption}
    % \label{fig:my_label}
\end{figure}
   
\subsection{Metrics}
We define the following three metrics to measure the rareness of the hardware designs.

\noindent \textbf{Definition 4:} \textit{Rarest Rareness} in a Design ($\Omega$)

\noindent In the following equation, $\Omega$ represents the rareness of the \textit{rarest} signal in the design. Consider an example design that has only two rare signals: $S^i$ and $S^j$ where $S^i_\omega$= 0.1 and $S^j_\omega$= 0.05. %If $S^j$ is `0' 5\% of the time (`1' for 95\% of the time), 
Then $\Omega$ = 0.05 (smallest between $S_\omega^i$ and $S_\omega^j$).
\vspace{-0.08in}
\begin{equation}\footnotesize
   \Omega = \min(\{S_{\omega}^0, .. , S_{\omega}^n\})
\end{equation}
\vspace{-0.2in}

%\vspace{0.05in}
\noindent \textbf{Definition 5}: \textit{Average Rareness} ($\mu({\omega_n})$)

\noindent We define average rareness for most rare $n$ signals as below. Clearly, \textit{higher average rareness implies that the design is more resistant against malicious implants}. In other words, higher average rareness implies easier Trojan detection.  
\vspace{-0.07in}
\begin{equation}\footnotesize
    \mu({\omega_n}) = \frac{\sum_{i=0}^n S_{\omega}^i}{n} 
\end{equation}
\vspace{-0.2in}

\noindent \textbf{Definition 6:} \textit{Signal Count} less than a threshold ($ \rho_{(<\tau)}$)

\noindent We count the number of rare signals with rareness less than a specific threshold $\tau$ in a design $D$ as follows. Clearly, \textit{lower $\rho_{(<\tau)}$ implies that the design is more resistant against malicious implants}. In other words, lower $\rho_{(<\tau)}$ indirectly implies easier Trojan detection for the given rareness threshold.
\vspace{-0.07in}
\begin{equation}\footnotesize
    \rho_{(<\tau)} = |\forall S^i_\omega \in D : S_{\omega}^i \leq \tau| 
\end{equation}
\vspace{-0.2in}

\subsection{Theoretical Analysis}\label{subsec:theory}
Rareness of the signals in a hardware design depends on the type and the order of logic gates involved in the propagation path of the considered signal. In this section, we first show how to compute the logic probability vector of a signal. Next, we analyze the effects of various parameters on rareness, including the types of logic gates in a specific path, logic depth (number of logic gates in a path), as well as design area (total number of logic gates).

\vspace{0.05in}
\subsubsection{\textbf{Calculating the Logic Probability Vector of a Signal}}\label{subsubsec:logicProbability}
\hfill\\
We formulate the rareness calculation of a signal as a matrix multiplication problem. Figure~\ref{fig:andEx} illustrates the example calculation for a fan-out of an AND gate. For this example we have to use the ITM corresponding to AND gate from Equation~\ref{eqn:itmgates}. Then we obtain Equation~\ref{eqn:interX} by multiplying the Kronecker product of input probabilities ($P(A)$ and $P(B)$) of the gate with the ITM matrix of the AND gate. The column sum of the resultant matrix represents the P(0) and P(1) values of the fan-out ($Z$) signal.

\begin{figure}[H]
\vspace{-0.25in}
\centering
\begin{subfigure}{.5\linewidth}
  \centering
  
 \begin{circuitikz}[line width=0.15pt]
 \draw
(0,1) node[and port] (and1) {}
    (and1.in 1) node [anchor=east] {A}
    (and1.in 2) node [anchor=east] {B}
    (and1.out) node [anchor=west] {Z};
\end{circuitikz}
\vspace{-0.1in}
    % \footnotesize
    %  $
    %     P(A)=P(B)=\begin{bmatrix}
    %     0.5 & 0 \\
    %     0 & 0.5 
    %     \end{bmatrix}
    %  $

\hrulefill\par
\end{subfigure}%
\begin{subfigure}{.5\linewidth}
  \centering
   
   \footnotesize
   \begin{equation*}
        P(A)=P(B)=\begin{bmatrix}
        0.5 & 0 \\
        0 & 0.5 
        \end{bmatrix}
        % ITM_{AND} =  \begin{bmatrix}
        % 1 & 0 \\
        % 1 & 0 \\
        % 1 & 0 \\
        % 0 & 1
        % \end{bmatrix} 
        \label{eqn:itmand}
   \end{equation*}

  \label{fig:andABZ}
  \hrulefill\par
\end{subfigure}

\begin{subfigure}{1\linewidth}
\vspace{-0.1in}
  \centering
    \begin{equation}
        X = P(A)\otimes P(B) \times ITM_{AND}
        \label{eqn:interX}
    \end{equation}
% \hrulefill
\end{subfigure}

\begin{subfigure}{1\linewidth}
  \centering
  \footnotesize
  \begin{equation*}
        \begin{bmatrix}
        a_1 & b_1 \\
        a_2 & b_2 \\
        a_3 & b_3 \\
        a_4 & b_4
        \end{bmatrix} = \begin{bmatrix}
        0.25 & 0 & 0 & 0\\
        0 & 0.25 & 0 & 0 \\
        0 & 0 & 0.25 & 0 \\
        0 & 0 & 0 & 0.25
        \end{bmatrix} \times \begin{bmatrix}
        1 & 0 \\
        1 & 0 \\
        1 & 0 \\
        0 & 1
        \end{bmatrix}
    \end{equation*}
    \hrulefill\par
    \vspace{-0.05in}
\end{subfigure}

\begin{subfigure}{1\linewidth}
  \centering
  \begin{equation}
        P(\bar Z) = < \sum_{i=1}^4 a_i ,  \sum_{i=1}^4 b_i > = <0.75, 0.25>
    \end{equation}
\end{subfigure}
% \vspace{-0.1in}
\caption{Calculating the rareness probability for a signal ($Z$) when fan-in signals ($A,B$) propagate through an AND gate.}
\label{fig:andEx}
%\vspace{-0.1in}
\end{figure}

\subsubsection{\textbf{Effect of Logic Gate Types in the Path on Rareness}}
\hfill\\
The types of logic gates involved in a circuit is a signature of the design. Since the ITM matrix (discussed above) is calculated based on the truth table of the logic gate, the rareness of a signal is affected by the type of logic involved in the circuit. In other words, P(1) is a lower value for the output of an AND gate, while an OR gate makes P(0) a lower value. Therefore, it is possible to obtain designs with lower rareness metric using different logic implementations. We will explore design diversity in Section~\ref{sec:reduction} to verify this hypothesis.

\vspace{0.05in}
\subsubsection{\textbf{Effect of Logic Depth on Rareness}}
\label{subsubsec:logicDepth}
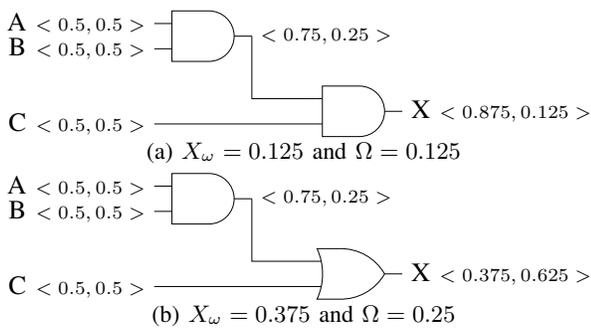
\begin{figure}[htp]
    \centering
    \vspace{-0.1in}
    \begin{subfigure}{1\linewidth}
  \centering
\begin{circuitikz}[line width=0.15pt]
\draw
(0,1) node[and port] (myand1) {}
    (myand1.in 1) node [anchor=east] {A \scriptsize{$<0.5,0.5>$}}
    (myand1.in 2) node [anchor=east] {B \scriptsize{$<0.5,0.5>$}}
    (myand1.out) node [anchor=west] { \scriptsize{$<0.75,0.25>$}}
(2,0) node[and port] (myand2) {}
    %(myand1.in 2) node [anchor=east] {C}
    (myand2.out) node [anchor=west] {X \scriptsize{$<0.875,0.125>$}}
    (myand1.out) |- (myand2.in 1);

\draw (myand2.in 2) to (myand2.in 2 -| myand1.in 2)
        node [anchor=east] {C \scriptsize{$<0.5,0.5>$}};

\end{circuitikz}
\vspace{-0.1in}
\caption{$X_{\omega} = 0.125$ and $\Omega = 0.125$}
    \label{fig:logic_prop_and}
\end{subfigure}
\begin{subfigure}{1\linewidth}
  \centering
\begin{circuitikz}[line width=0.15pt]
\draw
(0,1) node[and port] (myand1) {}
    (myand1.in 1) node [anchor=east] {A \scriptsize{$<0.5,0.5>$}}
    (myand1.in 2) node [anchor=east] {B \scriptsize{$<0.5,0.5>$}}
    (myand1.out) node [anchor=west] { \scriptsize{$<0.75,0.25>$}}
(2,0) node[or port] (myand2) {}
    %(myand1.in 2) node [anchor=east] {C}
    (myand2.out) node [anchor=west] {X \scriptsize{$<0.375,0.625>$}}
    (myand1.out) |- (myand2.in 1);

\draw (myand2.in 2) to (myand2.in 2 -| myand1.in 2)
        node [anchor=east] {C \scriptsize{$<0.5,0.5>$}};

\end{circuitikz}
\vspace{-0.1in}
\caption{$X_{\omega} = 0.375$ and $\Omega=0.25$}
    \label{fig:logic_prop_and_or}
\end{subfigure}
    % \vspace{-0.2in}
    \caption{An example to illustrate the effect of gate type in rareness propagation through logic depth.}
    \label{fig:logic_depth}
        \vspace{-0.05in}
\end{figure}

\hfill\\
Since the logic probability of a signal is always less than 1 ($P(\overline{signal}) \leq 1$), signal propagation through the same gate type will always reduce the 
probability. However, this phenomenon may not hold when gate types are interchanged in the propagation path. Figure~\ref{fig:logic_depth} illustrates a counter-example to demonstrate this scenario. Figure~\ref{fig:logic_prop_and} shows a circuit with a logic depth of two, with two AND gates. Figure~\ref{fig:logic_prop_and_or} consists of a similar circuit except the last AND gate is replaced by an OR gate. In  \ref{fig:logic_prop_and}, the fan-out signal is the rarest ($\Omega$) signal, although in \ref{fig:logic_prop_and_or} the rarest ($\Omega$) signal is not the  fan-out signal. This demonstrates that the effect of logic depth on the rareness values depends on the design. Therefore, we will explore rareness reduction techniques in Section~\ref{sec:reduction}.

\subsubsection{\textbf{Effect of Area Optimization on Rareness}}
\hfill\\
Logic optimization refers to reducing a complex logical equation to a simplified version without changing the indented behavior of the circuit. There are various logic optimization techniques for Boolean circuits such as Boolean algebra, graphical methods (e.g., Karnaugh maps, Quine–McCluskey algorithm, and Petrick's method), heuristic methods (e.g., Espresso heuristic logic minimizer), etc. During these optimizations, either logic gates get removed by gate sharing or a part of the circuit may get replaced with a simpler circuit. For example, Karnaugh map tries to identify repetitive patterns in the signals and eliminates them. Based on the intuition provided by logic optimization, we analyzed the relationship of rareness metrics ($\mu(\omega_n)$, $\rho_{(<r)}$) with logic area optimization. Results revealed that if the area reduction is occurred within the region that contributed towards the rareness metrics, then area optimization improves the rareness metrics. We have performed empirical analysis on real-world hardware designs with different synthesis area efforts, as demonstrated in Section~\ref{subsec:exp:area}.  
% Due to the elimination of logic gates combined with the corresponding nets, signal density though the remaining nets gets increased due to gate sharing while reducing the possibility of rare signals. This directly affects the two major metrics of average rareness ($\mu(\omega_n)$) and signal count less than a threshold rareness ($\rho_{(<r)}$). Since logic optimization gets rid of the sparse circuit if possible, it affects $\rho_{(<r)}$. Since $\mu(\omega_n)$ is interested in the minimum probability of the two possibilities of the signal being ``0" or ``1", removing elements from the average calculation bias the $\mu(\omega_n)$ value toward $0.5$ increasing average rareness.

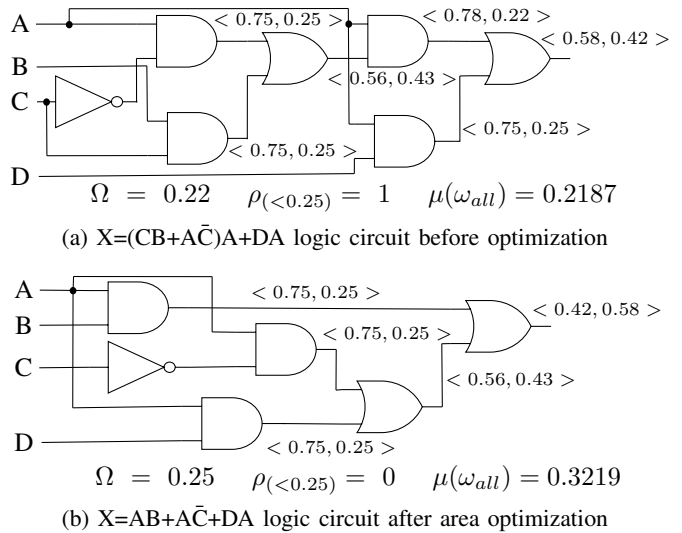
\begin{figure}[hbtp]
    \centering
    
    \begin{subfigure}{1\linewidth}
    \centering
    \input{Images/beforeOptEx}
    \vspace{-0.1in}
    \caption{X=(CB+A\=C)A+DA logic circuit before optimization}
    \vspace{0.1in}
    \label{subfig:unoptimizedLogic}
  
    % \hrulefill\par
\end{subfigure}
    \vspace{0.3in}
     \begin{subfigure}{1\linewidth}
    \centering
    \input{Images/afterOptEx}
    \vspace{-0.2in}
    \caption{X=AB+A\=C+DA logic circuit after area optimization}
    \label{subfig:optimizedLogic}
\end{subfigure}
     \vspace{-0.50in}
         \caption{An example scenario of rareness reduction  of logic circuits through area optimization}
    \label{fig:logicEx1}
    \vspace{-0.1in}
\end{figure}

Figure~\ref{fig:logicEx1} presents an illustrative example to demonstrate the effect of area reduction on rareness metrics. Figure~\ref{subfig:unoptimizedLogic} shows the logic circuit representing the Boolean equation \textit{X=(CB+A\=C)A+DA}. Corresponding $\Omega, \rho_{(<0.25)}, $ and $\mu(\omega_{all})$ metric values are 0.2187, 1, and 0.3033 respectively. Then using  Karnaugh maps we optimized the circuit \textit{(CB+A\=C)A+DA } to equivalent Boolean circuit \textit{(AB+A\=C)+DA}, which is represented in Figure~\ref{subfig:optimizedLogic}. From the metric values calculated for the optimized circuit, it can be observed that the rarest signal ($\Omega$) has increased from $0.2187$ to $0.25$. The number of signals less than the threshold of $0.25$ has reduced from $1$ to $0$ and the average rareness of all the signals has increased from $0.3033$ to $0.3218$ demonstrating that area reduction has positively impacted rareness reduction. Table~\ref{tab:booleanEx} presents several illustrative examples for different logic circuit expressions.

\begin{table}[hbtp]
\centering
% \vspace{-0.1in}
\caption{Rareness metrics for different Boolean circuit expressions before and after area optimizations}
\label{tab:booleanEx}
\vspace{-0.09in}
\begin{tabular}{|llll|llll|}
\hline
\multicolumn{4}{|c|}{\textbf{Original Circuit}}                                                    & \multicolumn{4}{c|}{\textbf{Area Optimized Circuit}}                                              \\ \hline
\multicolumn{1}{|l|}{\textit{Logic}}      & \multicolumn{1}{l|}{$\Omega$}  & \multicolumn{1}{l|}{$\rho$} & $\mu$     & \multicolumn{1}{l|}{\textit{Logic}}  & \multicolumn{1}{l|}{$\Omega$} & \multicolumn{1}{l|}{$\rho$} &$\mu$      \\ \hline
\multicolumn{1}{|l|}{\textit{AB+BC(B+C)}} & \multicolumn{1}{l|}{0.18} & \multicolumn{1}{l|}{1}   & 0.26 & \multicolumn{1}{l|}{\textit{(A+C)B}} & \multicolumn{1}{l|}{0.25}  & \multicolumn{1}{l|}{0}   & 0.31 \\ \hline
\multicolumn{1}{|l|}{\textit{AC+A\=B\=C+ABC}}           & \multicolumn{1}{l|}{0.12}       & \multicolumn{1}{l|}{2}    &    0.25    & \multicolumn{1}{l|}{\textit{A\=B+AC}}       & \multicolumn{1}{l|}{0.25}      & \multicolumn{1}{l|}{0}    &    0.31    \\ \hline
\multicolumn{1}{|l|}{\textit{ADC+ABD}}           & \multicolumn{1}{l|}{0.12}       & \multicolumn{1}{l|}{2}    &    0.19    & \multicolumn{1}{l|}{\textit{A(DC+BD)}}       & \multicolumn{1}{l|}{0.21}      & \multicolumn{1}{l|}{0}    &    0.28    \\ \hline
\end{tabular}
\vspace{-0.2in}
\end{table}

\subsection{Rareness Reduction Techniques}
\label{sec:reduction}

Based on the theoretical analysis in Section~\ref{subsec:theory}, we can conclude that two factors affect the rareness metrics in a hardware design. (i) nature of the design and (ii) area of the design. We propose two techniques to reduce the rareness of signals in hardware designs considering the above factors.

\vspace{0.05in}
\subsubsection{\textbf{Design Diversity}} In order to achieve a functionality, there can be multiple algorithms. There are multiple readily available implementations for most generic sub-components, such as adders, multipliers, dividers, sorting algorithms, search algorithms, hashing algorithms, etc. We explore different implementations for sub-components of the design. For example, if we need an adder, we can consider various adder choices (e.g., ripple-carry adder, carry lookahead adder, etc.) to select the implementation with the minimum contribution to the rareness. Similarly, if we need to implement sorting, we can consider diverse sorting algorithms, including bubble sort, insertion sort, quick sort, merge sort, etc. while we are trying to improve rareness, we also have to satisfy other design constraints, such as area, power, and performance.

\vspace{0.05in}
\subsubsection{ \textbf{Area Optimization}} Our theoretical analysis revealed that area reduction leads to improved rareness. Therefore, a design-for-trust solution needs to select the implementation with the lowest area without violating other design constraints. Another way to reduce design area is by reducing the parallelism inside the design. Any hardware synthesis tool considers various avenues for area reduction including parallelism reduction (sharing components), simplified (bare-bone) implementation, and logic minimization. %Avoiding redundant components that significantly contribute towards rareness as much as possible will improve the rareness metrics. 
For example, let us consider a processor consisting of two ALU units. 
%which has a $\rho_{(<0.1)}$ of $x$ in each. 
If we use a single ALU, it is expected  to reduce the $\rho_{(<0.1)}$ value contribution of ALU's by $\frac{1}{2}$. Similarly, the bare-bone implementation of the required functionality is preferable for obtaining a verification-friendly design-for-trust solution. The logic minimization techniques are expected to reduce the area and improve the rareness. %Therefore, if the priority is Design-for-Trust we suggest selecting the bare-bone implementation technique during the design time.

\subsection{Fast Detection of Trojans with Rareness Reduction}

Existing Trojan detection techniques (\cite{mero2009swarup,Mers2016yuanwen,pan2021TGRL,cruz2018ATPG,tarmac2021yangdi,jayasena2023scalable}) follows the threat model outlined in Section~\ref{sec:threat}. In other words, the Trojan detection time depends on the number of rare signals in the design.
To evaluate the effects of rareness reduction on Trojan detection, we consider two complimentary test generation based Trojan detection techniques:  statistical~\cite{mero2009swarup} and maximal clique activation~\cite{tarmac2021yangdi}.

\vspace{0.05in}
\subsubsection{\textbf{Trojan Detection using Statistical Test Generation}}
\hfill\\
Statistical test generation technique MERO~\cite{mero2009swarup} depends on N-detect~\cite{pomeranz2004measure} principle, where each rare signal is activated $N$ times. First, it simulates the design with random test vectors while performing rareness calculations. Next, it identifies all the rare signals (potential trigger conditions) with rareness values less than a specific threshold ($\tau$). Then using the initial set of random test vectors, the algorithm performs bit flips until the $N$ criterion is satisfied for all the identified rare signals. Due to the statistical nature of the generated test set, if $N$ is sufficiently large, a good Trojan coverage can be obtained. The authors demonstrate the results of the MERO framework on ISCAS'85 benchmarks. To achieve a good coverage of detecting Trojan triggers consisting of four triggers, the authors have used a $N$ value of 1000.
% During the rareness analysis, authors found 724 average rare signals for the rareness threshold($\tau$) of 0.2 out of the total average signal count of 2857. 
Rareness reduction is effective for statistical-based test generation in two ways. (1) It reduces the number of rare signals in the design. Assume that the number of rare signals that we can reduce is $X$. This reduces the initial rareness calculation time by reducing the signal value monitoring effort by $X$. This further reduces the test generation in the order of $X\times N$, (ii) Reducing the average rareness of the design improves the chances of signals getting activated during random simulations as well as during the execution of the underlying bit-flipping algorithm, yielding higher Trojan coverage from the generated test vectors. Section~\ref{subsec:exp:testgeneration} demonstrates the effect of rareness reduction on statistical test generation.

\vspace{0.05in}
\subsubsection{\textbf{Trojan detection  using Maximal Clique Activation}}
Directed test generation technique of 
TARMAC~\cite{tarmac2021yangdi} tackles the problem following a complementary approach to MERO. Similar to MERO, TARMAC first calculates the rare signals in the system with random test vectors. Let us assume that we have identified $R$ number of rare signals. For all the rare signals, TARMAC creates a two-trigger connectivity graph by querying all pairs ($R \times \frac{R-1}{2}$) of rare signals using satisfiability solving. The complexity of the satisfiability graph construction is in the order of $R^2$. Next, maximal clique partitioning is employed on the satisfiability graph to identify the trigger cliques. SAT solver is used to generate test vectors to activate   all the identified cliques in the design. Rareness reduction benefits TARMAC in two ways. (1) Suppose the number of rare signals that we can reduce is $X$. Then satisfiability graph construction complexity is reduced in the order of ${(R-X)^2}$. (2) Due to the reduction of average rareness, it is easier for the SAT solvers to activate the cliques. This significantly reduces the three major limitations of TARMAC, satisfiability graph construction, clique partitioning, and test generation using clique activation. Section~\ref{subsec:exp:testgeneration} demonstrates the effect of rareness reduction on maximal clique activation.

%% file: Images/overview.tex
\tikzset{every picture/.style={line width=0.75pt}} %set default line width to 0.75pt        
% \footnotesize
\begin{tikzpicture}[x=0.75pt,y=0.75pt,yscale=-0.85,xscale=0.95]
%uncomment if require: \path (0,419); %set diagram left start at 0, and has height of 419

%Shape: Rectangle [id:dp6952430975237776] 
\draw  [draw opacity=0][fill={rgb, 255:red, 153; green, 0; blue, 77 }  ,fill opacity=1 ][line width=0.75]  (-1.75,43.85) -- (107.26,43.85) -- (107.26,83.3) -- (-1.75,83.3) -- cycle ;

%Shape: Rectangle [id:dp881435087966637] 
\draw  [draw opacity=0][fill={rgb, 255:red, 153; green, 0; blue, 77 }  ,fill opacity=1 ][line width=0.75]  (620.31,43.97) -- (703.4,43.97) -- (703.4,83.42) -- (620.31,83.42) -- cycle ;

%Shape: Ellipse [id:dp44612950755679925] 
\draw  [draw opacity=0][fill={rgb, 255:red, 120; green, 89; blue, 163 }  ,fill opacity=1 ][line width=0.75]  (529.96,63.28) .. controls (529.96,52.16) and (547.43,43.14) .. (568.97,43.14) .. controls (590.52,43.14) and (607.98,52.16) .. (607.98,63.28) .. controls (607.98,74.41) and (590.52,83.42) .. (568.97,83.42) .. controls (547.43,83.42) and (529.96,74.41) .. (529.96,63.28) -- cycle ;
%Straight Lines [id:da7503256845371262] 
\draw [line width=0.75]    (607.98,63.28) -- (617.09,63.39) ;
\draw [shift={(620.09,63.42)}, rotate = 180.66] [fill={rgb, 255:red, 0; green, 0; blue, 0 }  ][line width=0.08]  [draw opacity=0] (8.93,-4.29) -- (0,0) -- (8.93,4.29) -- cycle    ;
%Flowchart: Alternative Process [id:dp5431944453406603] 
\draw   (523.76,39.9) .. controls (523.76,34.11) and (528.45,29.42) .. (534.24,29.42) -- (697.81,29.42) .. controls (703.6,29.42) and (708.29,34.11) .. (708.29,39.9) -- (708.29,78.83) .. controls (708.29,84.61) and (703.6,89.31) .. (697.81,89.31) -- (534.24,89.31) .. controls (528.45,89.31) and (523.76,84.61) .. (523.76,78.83) -- cycle ;
%Shape: Rectangle [id:dp28950682174484643] 
\draw  [draw opacity=0][fill={rgb, 255:red, 255; green, 255; blue, 255 }  ,fill opacity=1 ] (553.67,17.92) -- (679.67,17.92) -- (679.67,41.81) -- (553.67,41.81) -- cycle ;
%Shape: Rectangle [id:dp46870860810889914] 
\draw  [draw opacity=0][fill={rgb, 255:red, 153; green, 0; blue, 77 }  ,fill opacity=1 ][line width=0.75]  (395.39,43.97) -- (504.4,43.97) -- (504.4,83.42) -- (395.39,83.42) -- cycle ;

%Shape: Rectangle [id:dp677151317452948] 
\draw  [draw opacity=0][fill={rgb, 255:red, 120; green, 89; blue, 163 }  ,fill opacity=1 ][line width=0.75]  (136.3,44.35) -- (245.3,44.35) -- (245.3,83.8) -- (136.3,83.8) -- cycle ;

%Straight Lines [id:da8146553226385576] 
\draw [line width=0.75]    (106.71,62.95) -- (124.18,63.03) ;
\draw [shift={(127.18,63.04)}, rotate = 180.25] [fill={rgb, 255:red, 0; green, 0; blue, 0 }  ][line width=0.08]  [draw opacity=0] (8.93,-4.29) -- (0,0) -- (8.93,4.29) -- cycle    ;
%Flowchart: Alternative Process [id:dp9126177476335895] 
\draw   (127.55,39.78) .. controls (127.55,33.94) and (132.28,29.21) .. (138.12,29.21) -- (362.89,29.21) .. controls (368.73,29.21) and (373.46,33.94) .. (373.46,39.78) -- (373.46,79.04) .. controls (373.46,84.88) and (368.73,89.61) .. (362.89,89.61) -- (138.12,89.61) .. controls (132.28,89.61) and (127.55,84.88) .. (127.55,79.04) -- cycle ;
%Shape: Rectangle [id:dp4982309772678484] 
\draw  [draw opacity=0][fill={rgb, 255:red, 120; green, 89; blue, 163 }  ,fill opacity=1 ][line width=0.75]  (255.34,44.18) -- (364.34,44.18) -- (364.34,83.63) -- (255.34,83.63) -- cycle ;

%Straight Lines [id:da1308799175558617] 
\draw [line width=0.75]    (373.99,63.29) -- (391.45,63.36) ;
\draw [shift={(394.45,63.38)}, rotate = 180.25] [fill={rgb, 255:red, 0; green, 0; blue, 0 }  ][line width=0.08]  [draw opacity=0] (8.93,-4.29) -- (0,0) -- (8.93,4.29) -- cycle    ;
%Straight Lines [id:da6997395175823364] 
\draw [line width=0.75]    (504.48,62.89) -- (521.94,62.96) ;
\draw [shift={(524.94,62.98)}, rotate = 180.25] [fill={rgb, 255:red, 0; green, 0; blue, 0 }  ][line width=0.08]  [draw opacity=0] (8.93,-4.29) -- (0,0) -- (8.93,4.29) -- cycle    ;
%Shape: Rectangle [id:dp31081956610781214] 
\draw  [draw opacity=0][fill={rgb, 255:red, 255; green, 255; blue, 255 }  ,fill opacity=1 ] (186.4,20.12) -- (313.2,20.12) -- (313.2,39.14) -- (186.4,39.14) -- cycle ;

% Text Node
\draw (52.75,63.7) node   [align=left] {\begin{minipage}[lt]{74.12pt}\setlength\topsep{0pt}
\begin{center}
\textcolor[rgb]{1,1,1}{\textbf{Design}}\\\textcolor[rgb]{1,1,1}{{\small (RTL/GateLevel)}}
\end{center}

\end{minipage}};
% Text Node
\draw (661.86,63.66) node   [align=left] {\begin{minipage}[lt]{56.5pt}\setlength\topsep{0pt}
\begin{center}
\textbf{\textcolor[rgb]{1,1,1}{Effective}}\\\textcolor[rgb]{1,1,1}{\textbf{Detection}}
\end{center}

\end{minipage}};
% Text Node
\draw (449.89,63.66) node   [align=left] {\begin{minipage}[lt]{74.12pt}\setlength\topsep{0pt}
\begin{center}
\textcolor[rgb]{1,1,1}{\textbf{Netlist}}\\\textcolor[rgb]{1,1,1}{{\small (GateLevel)}}
\end{center}

\end{minipage}};
% Text Node
\draw (190.8,64.03) node   [align=left] {\begin{minipage}[lt]{74.12pt}\setlength\topsep{0pt}
\begin{center}
\textcolor[rgb]{1,1,1}{\textbf{Design}}\\\textcolor[rgb]{1,1,1}{\textbf{Diversity}}
\end{center}

\end{minipage}};
% Text Node
\draw (309.84,63.86) node   [align=left] {\begin{minipage}[lt]{74.12pt}\setlength\topsep{0pt}
\begin{center}
\textcolor[rgb]{1,1,1}{\textbf{Area}}\\\textcolor[rgb]{1,1,1}{\textbf{Optimization}}
\end{center}

\end{minipage}};
% Text Node
\draw (616.49,29.78) node   [align=left] {\begin{minipage}[lt]{112.36pt}\setlength\topsep{0pt}
\begin{center}
Fast Trojan Detection
\end{center}

\end{minipage}};
% Text Node
\draw (250.5,28.43) node   [align=left] {\begin{minipage}[lt]{102pt}\setlength\topsep{0pt}
\begin{center}
Rareness Reduction
\end{center}

\end{minipage}};
% Text Node
\draw (568.5,61.35) node   [align=left] {\begin{minipage}[lt]{53.05pt}\setlength\topsep{0pt}
\begin{center}
\textbf{\textcolor[rgb]{1,1,1}{Framework}}
\end{center}

\end{minipage}};

\end{tikzpicture}

%% file: Images/beforeOptEx.tex
% \tikzset{every picture/.style={line width=0.75pt}} %set default line width to 0.75pt        

\begin{tikzpicture}[x=0.75pt,y=0.75pt,yscale=-.9,xscale=0.8]

%Shape: And Gate [id:dp627545974038036] 
\draw   (280.03,80.91) -- (299.02,80.91) .. controls (309.5,80.91) and (318.01,87.4) .. (318.01,95.38) .. controls (318.01,103.37) and (309.5,109.85) .. (299.02,109.85) -- (280.03,109.85) -- (280.03,80.91) -- cycle (267.37,85.74) -- (280.03,85.74) (267.37,105.03) -- (280.03,105.03) (318.01,95.38) -- (330.67,95.38) ;
%Shape: And Gate [id:dp4843918513208536] 
\draw   (149.53,79.27) -- (168.52,79.27) .. controls (179,79.27) and (187.51,85.75) .. (187.51,93.74) .. controls (187.51,101.72) and (179,108.21) .. (168.52,108.21) -- (149.53,108.21) -- (149.53,79.27) -- cycle (136.87,84.09) -- (149.53,84.09) (136.87,103.38) -- (149.53,103.38) (187.51,93.74) -- (200.17,93.74) ;
%Straight Lines [id:da24000602035480578] 
\draw    (326.5,39.25) -- (340.18,39.34) ;
%Straight Lines [id:da30234853850358556] 
\draw    (66.63,29.67) -- (130,29.64) ;
%Straight Lines [id:da2636795284179676] 
\draw    (87.58,21.04) -- (171.6,20.99) -- (263.55,20.91) ;
%Straight Lines [id:da6912676981025552] 
\draw    (87.58,21.04) -- (87.6,30.47) ;
%Straight Lines [id:da9150838969306017] 
\draw    (263.55,20.91) -- (264,85.72) ;
%Straight Lines [id:da6707169939767097] 
\draw    (199.86,94.21) -- (199.91,58.55) ;
%Shape: Ellipse [id:dp9777493158963781] 
\draw  [fill={rgb, 255:red, 0; green, 0; blue, 0 }  ,fill opacity=1 ][line width=0.75]  (86.1,29.97) .. controls (86.1,29.14) and (86.77,28.47) .. (87.6,28.47) .. controls (88.43,28.47) and (89.1,29.14) .. (89.1,29.97) .. controls (89.1,30.8) and (88.43,31.47) .. (87.6,31.47) .. controls (86.77,31.47) and (86.1,30.8) .. (86.1,29.97) -- cycle ;
%Shape: Or Gate [id:dp20227770609480844] 
\draw   (349.67,34.51) -- (365.49,34.51) .. controls (376.53,34.77) and (386.4,40.41) .. (390.81,48.98) .. controls (386.4,57.56) and (376.53,63.19) .. (365.49,63.45) -- (349.67,63.45) .. controls (356.45,54.5) and (356.45,43.47) .. (349.67,34.51) -- cycle (340.18,39.34) -- (352.83,39.34) (340.18,58.63) -- (352.83,58.63) (390.81,48.98) -- (403.47,48.98) ;
%Straight Lines [id:da1597754812734189] 
\draw    (333.83,58.75) -- (340.18,58.63) ;
%Straight Lines [id:da8746421857571971] 
\draw    (136.87,84.09) -- (136.54,53.09) ;
%Straight Lines [id:da8324759687419554] 
\draw    (67.1,53.71) -- (136.54,53.09) ;
%Shape: Not/Inverter Gate [id:dp2127676291899887] 
\draw   (79.03,58.74) -- (114.19,73.22) -- (79.03,87.69) -- (79.03,58.74) -- cycle (67.31,73.22) -- (79.03,73.22) (121.22,73.22) -- (130.6,73.22) (114.19,73.22) .. controls (114.19,71.62) and (115.76,70.32) .. (117.71,70.32) .. controls (119.65,70.32) and (121.22,71.62) .. (121.22,73.22) .. controls (121.22,74.81) and (119.65,76.11) .. (117.71,76.11) .. controls (115.76,76.11) and (114.19,74.81) .. (114.19,73.22) -- cycle ;
%Straight Lines [id:da9194002189750914] 
\draw    (68.4,115.1) -- (267.33,113.5) ;
%Straight Lines [id:da43925364936992817] 
\draw    (73.53,103.47) -- (136.87,103.38) ;
%Straight Lines [id:da7096220471552968] 
\draw    (73.53,103.47) -- (73.32,73.37) ;
%Straight Lines [id:da9259503077779638] 
\draw    (130.05,73.52) -- (130,48.94) ;
%Straight Lines [id:da4742261328575532] 
\draw    (333.83,58.75) -- (333.89,95.39) -- (330.67,95.38) ;
%Straight Lines [id:da1289586673643961] 
\draw    (264,85.72) -- (267.37,85.74) ;
%Shape: And Gate [id:dp10575582051499688] 
\draw   (275.86,24.78) -- (294.85,24.78) .. controls (305.33,24.78) and (313.84,31.26) .. (313.84,39.25) .. controls (313.84,47.24) and (305.33,53.72) .. (294.85,53.72) -- (275.86,53.72) -- (275.86,24.78) -- cycle (263.21,29.6) -- (275.86,29.6) (263.21,48.9) -- (275.86,48.9) (313.84,39.25) -- (326.5,39.25) ;
%Shape: Or Gate [id:dp011256001606565258] 
\draw   (209.41,34.43) -- (225.23,34.43) .. controls (236.27,34.69) and (246.13,40.33) .. (250.55,48.9) .. controls (246.13,57.47) and (236.27,63.11) .. (225.23,63.37) -- (209.41,63.37) .. controls (216.19,54.41) and (216.19,43.38) .. (209.41,34.43) -- cycle (199.91,39.25) -- (212.57,39.25) (199.91,58.55) -- (212.57,58.55) (250.55,48.9) -- (263.21,48.9) ;
%Shape: And Gate [id:dp4558548469669792] 
\draw   (142.66,24.82) -- (161.65,24.82) .. controls (172.13,24.82) and (180.64,31.3) .. (180.64,39.29) .. controls (180.64,47.28) and (172.13,53.76) .. (161.65,53.76) -- (142.66,53.76) -- (142.66,24.82) -- cycle (130,29.64) -- (142.66,29.64) (130,48.94) -- (142.66,48.94) (180.64,39.29) -- (193.29,39.29) ;
%Shape: Ellipse [id:dp27245430407081606] 
\draw  [fill={rgb, 255:red, 0; green, 0; blue, 0 }  ,fill opacity=1 ][line width=0.75]  (71.82,73.37) .. controls (71.82,72.54) and (72.49,71.87) .. (73.32,71.87) .. controls (74.14,71.87) and (74.82,72.54) .. (74.82,73.37) .. controls (74.82,74.2) and (74.14,74.87) .. (73.32,74.87) .. controls (72.49,74.87) and (71.82,74.2) .. (71.82,73.37) -- cycle ;
%Straight Lines [id:da1983283636580444] 
\draw    (199.79,39.29) -- (193.29,39.29) ;
%Shape: Ellipse [id:dp2595729425762412] 
\draw  [fill={rgb, 255:red, 0; green, 0; blue, 0 }  ,fill opacity=1 ][line width=0.75]  (262.32,29.64) .. controls (262.32,28.81) and (262.99,28.14) .. (263.82,28.14) .. controls (264.65,28.14) and (265.32,28.81) .. (265.32,29.64) .. controls (265.32,30.46) and (264.65,31.14) .. (263.82,31.14) .. controls (262.99,31.14) and (262.32,30.46) .. (262.32,29.64) -- cycle ;
%Straight Lines [id:da10163559518417675] 
\draw    (267.37,105.03) -- (267.33,113.5) ;

% Text Node
\draw (61.26,29.12) node   [align=left] {\begin{minipage}[lt]{12.68pt}\setlength\topsep{0pt}
A
\end{minipage}};
% Text Node
\draw (61.52,72.93) node   [align=left] {\begin{minipage}[lt]{12.68pt}\setlength\topsep{0pt}
C
\end{minipage}};
% Text Node
\draw (61.78,53.5) node   [align=left] {\begin{minipage}[lt]{12.68pt}\setlength\topsep{0pt}
B
\end{minipage}};
% Text Node
\draw (61.26,115.15) node   [align=left] {\begin{minipage}[lt]{12.68pt}\setlength\topsep{0pt}
D
\end{minipage}};
% Text Node
\draw (185.82,95.33) node [anchor=north west][inner sep=0.75pt]  [font=\scriptsize]  {${\textstyle < 0.75,0.25> }$};
% Text Node
\draw (333.13,83.27) node [anchor=north west][inner sep=0.75pt]  [font=\scriptsize]  {${\textstyle < 0.75,0.25> }$};
% Text Node
\draw (248.44,55.05) node [anchor=north west][inner sep=0.75pt]  [font=\scriptsize]  {${\textstyle < 0.56,0.43> }$};
% Text Node
\draw (382.8,32.87) node [anchor=north west][inner sep=0.75pt]  [font=\scriptsize]  {${\textstyle < 0.58,0.42> }$};
% Text Node
\draw (308,19.87) node [anchor=north west][inner sep=0.75pt]  [font=\scriptsize]  {${\textstyle < 0.78,0.22> }$};
% Text Node
\draw (176.8,21.87) node [anchor=north west][inner sep=0.75pt]  [font=\scriptsize]  {${\textstyle < 0.75,0.25> }$};
% Text Node
\draw (90,111) node [anchor=north west][inner sep=0.75pt]   [align=left] {\begin{minipage}[lt]{210.15pt}\setlength\topsep{0pt}
\begin{center}
\vspace{-0.1in}
\begin{equation*}
\Omega \ =\ 0.22\ \ \ \ \rho _{( < 0.25)} =\ 1\ \ \ \ \mu ( \omega _{all}) =0.2187
\end{equation*}
\vspace{-0.3in}
\end{center}

\end{minipage}};

\end{tikzpicture}

%% file: Images/afterOptEx.tex
\begin{tikzpicture}[x=0.75pt,y=0.75pt,yscale=-.9,xscale=0.8]
%Shape: And Gate [id:dp627545974038036] 
\draw   (109.71,25.08) -- (128.7,25.08) .. controls (139.18,25.08) and (147.69,31.57) .. (147.69,39.55) .. controls (147.69,47.54) and (139.18,54.03) .. (128.7,54.03) -- (109.71,54.03) -- (109.71,25.08) -- cycle (97.06,29.91) -- (109.71,29.91) (97.06,49.2) -- (109.71,49.2) (147.69,39.55) -- (160.35,39.55) ;
%Shape: And Gate [id:dp4843918513208536] 
\draw   (168.78,90.02) -- (187.77,90.02) .. controls (198.25,90.02) and (206.76,96.5) .. (206.76,104.49) .. controls (206.76,112.47) and (198.25,118.96) .. (187.77,118.96) -- (168.78,118.96) -- (168.78,90.02) -- cycle (156.12,94.84) -- (168.78,94.84) (156.12,114.13) -- (168.78,114.13) (206.76,104.49) -- (219.42,104.49) ;
%Shape: And Gate [id:dp6169444626861784] 
\draw   (202.87,48.47) -- (221.86,48.47) .. controls (232.34,48.47) and (240.85,54.95) .. (240.85,62.94) .. controls (240.85,70.93) and (232.34,77.41) .. (221.86,77.41) -- (202.87,77.41) -- (202.87,48.47) -- cycle (190.21,53.29) -- (202.87,53.29) (190.21,72.59) -- (202.87,72.59) (240.85,62.94) -- (253.51,62.94) ;
%Shape: Or Gate [id:dp4967398852627647] 
\draw   (266.61,80.65) -- (282.44,80.65) .. controls (293.47,80.91) and (303.34,86.55) .. (307.75,95.12) .. controls (303.34,103.69) and (293.47,109.33) .. (282.44,109.59) -- (266.61,109.59) .. controls (273.4,100.64) and (273.4,89.6) .. (266.61,80.65) -- cycle (257.12,85.47) -- (269.78,85.47) (257.12,104.77) -- (269.78,104.77) (307.75,95.12) -- (320.41,95.12) ;
%Straight Lines [id:da1815388944823273] 
\draw    (257.12,104.77) -- (219.42,104.49) -- (221.82,104.47) ;
%Straight Lines [id:da24000602035480578] 
\draw    (160.35,39.55) -- (325.68,40.34) ;
%Straight Lines [id:da30234853850358556] 
\draw    (66.63,29.67) -- (97.06,29.91) ;
%Straight Lines [id:da2636795284179676] 
\draw    (87.93,22.03) -- (175.28,22.11) ;
%Straight Lines [id:da6912676981025552] 
\draw    (87.93,22.03) -- (87.6,30.47) ;
%Straight Lines [id:da9150838969306017] 
\draw    (253.51,62.94) -- (253.37,85.74) ;
%Straight Lines [id:da6707169939767097] 
\draw    (161.37,72.93) -- (190.21,72.59) ;
%Shape: Ellipse [id:dp9777493158963781] 
\draw  [fill={rgb, 255:red, 0; green, 0; blue, 0 }  ,fill opacity=1 ][line width=0.75]  (86.1,29.97) .. controls (86.1,29.14) and (86.77,28.47) .. (87.6,28.47) .. controls (88.43,28.47) and (89.1,29.14) .. (89.1,29.97) .. controls (89.1,30.8) and (88.43,31.47) .. (87.6,31.47) .. controls (86.77,31.47) and (86.1,30.8) .. (86.1,29.97) -- cycle ;
%Shape: Or Gate [id:dp20227770609480844] 
\draw   (335.17,35.51) -- (350.99,35.51) .. controls (362.03,35.77) and (371.9,41.41) .. (376.31,49.98) .. controls (371.9,58.56) and (362.03,64.19) .. (350.99,64.45) -- (335.17,64.45) .. controls (341.95,55.5) and (341.95,44.47) .. (335.17,35.51) -- cycle (325.68,40.34) -- (338.33,40.34) (325.68,59.63) -- (338.33,59.63) (376.31,49.98) -- (388.97,49.98) ;
%Straight Lines [id:da1597754812734189] 
\draw    (320.07,59.74) -- (325.68,59.63) ;
%Straight Lines [id:da5968688148479628] 
\draw    (320.07,59.74) -- (320.41,95.12) ;
%Straight Lines [id:da8746421857571971] 
\draw    (66.31,73.22) -- (98.08,72.93) ;
%Straight Lines [id:da8324759687419554] 
\draw    (66.48,49.36) -- (97.06,49.2) ;
%Shape: Not/Inverter Gate [id:dp2127676291899887] 
\draw   (109.8,58.46) -- (144.96,72.93) -- (109.8,87.4) -- (109.8,58.46) -- cycle (98.08,72.93) -- (109.8,72.93) (151.99,72.93) -- (161.37,72.93) (144.96,72.93) .. controls (144.96,71.33) and (146.53,70.03) .. (148.48,70.03) .. controls (150.42,70.03) and (151.99,71.33) .. (151.99,72.93) .. controls (151.99,74.53) and (150.42,75.82) .. (148.48,75.82) .. controls (146.53,75.82) and (144.96,74.53) .. (144.96,72.93) -- cycle ;
%Straight Lines [id:da9194002189750914] 
\draw    (67.78,115.11) -- (156.12,114.13) ;
%Straight Lines [id:da43925364936992817] 
\draw    (87.4,95.67) -- (156.12,94.84) ;
%Straight Lines [id:da7096220471552968] 
\draw    (87.4,95.67) -- (87.6,30.47) ;
%Straight Lines [id:da9259503077779638] 
\draw    (175.28,53.11) -- (175.28,22.11) ;
%Straight Lines [id:da4742261328575532] 
\draw    (190.21,53.29) -- (175.28,53.11) ;
%Straight Lines [id:da1289586673643961] 
\draw    (253.37,85.74) -- (257.12,85.47) ;

% Text Node
\draw (61.26,29.12) node   [align=left] {\begin{minipage}[lt]{12.68pt}\setlength\topsep{0pt}
A
\end{minipage}};
% Text Node
\draw (61.52,72.93) node   [align=left] {\begin{minipage}[lt]{12.68pt}\setlength\topsep{0pt}
C
\end{minipage}};
% Text Node
\draw (61.67,49.39) node   [align=left] {\begin{minipage}[lt]{12.68pt}\setlength\topsep{0pt}
B
\end{minipage}};
% Text Node
\draw (61.26,115.15) node   [align=left] {\begin{minipage}[lt]{12.68pt}\setlength\topsep{0pt}
D
\end{minipage}};
% Text Node
\draw (375.23,35.6) node [anchor=north west][inner sep=0.75pt]  [font=\scriptsize]  {${\textstyle < 0.42,0.58> }$};
% Text Node
\draw (100.01,126.6) node [anchor=north west][inner sep=0.75pt]    {$\Omega \ =\ 0.25\ \ \ \ \rho _{( < 0.25)} =\ 0\ \ \ \ \mu ( \omega _{all}) =0.3219$};
% Text Node
\draw (198,26.27) node [anchor=north west][inner sep=0.75pt]  [font=\scriptsize]  {${\textstyle < 0.75,0.25> }$};
% Text Node
\draw (242.8,49.07) node [anchor=north west][inner sep=0.75pt]  [font=\scriptsize]  {${\textstyle < 0.75,0.25> }$};
% Text Node
\draw (207.6,112.87) node [anchor=north west][inner sep=0.75pt]  [font=\scriptsize]  {${\textstyle < 0.75,0.25> }$};
% Text Node
\draw (320.8,74.67) node [anchor=north west][inner sep=0.75pt]  [font=\scriptsize]  {${\textstyle < 0.56,0.43> }$};

\end{tikzpicture}

%% file: sections/experiments.tex
\section{Experiments} \label{sec:experiments}
We have created several experimental scenarios to strengthen the analysis of rareness reduction techniques in Section~\ref{sec:methodology}. First, we explain the experimental setup. Next, we conduct rareness reduction experiments using design diversity as well as area optimization. Finally, we evaluate the effects of rareness reduction on detecting randomly inserted Trojans.

\subsection{Experimental Setup}
\begin{figure}[htp]
    \begin{center}
    \vspace{-0.1in}
    \begin{adjustwidth*}{-1em}{-1em}
        \footnotesize
        \input{Images/empExperiments}
    \end{adjustwidth*}
    \end{center}
      \vspace{-0.1in}
      \caption{Overview of our evaluation framework.}
        \vspace{-0.2in}
      \label{fig:exp1}
\end{figure}
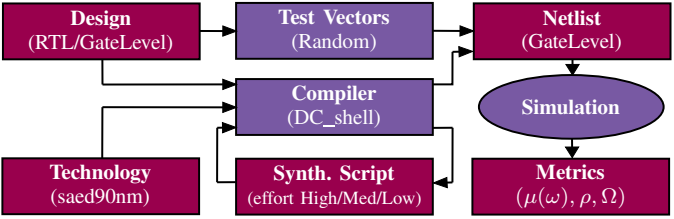

All the experiments including the execution of state-of-the-art test generation methods were carried out on a server with Intel(R) Xeon(R) CPU E5-2640 v3 @2.60GHz processor and 64GiB Memory.
For rareness and coverage analysis simulations, we have used \textit{Synopsys VCS} simulator. For compiling the RTL designs to the gate-level netlist, \textit{Synopsys DC Compiler} is used with \textit{SAED90nm} CMOS technology. In order to calculate the rareness of the synthesized designs, we have obtained the VCD dump of the synthesized designs. For validating the sampled Trojan triggers, \textit{Synopsys TetraMax}  was used. An overview of the experimental setup used for the evaluation is presented in Figure~\ref{fig:exp1}.

\subsection{Design Diversity Experiment}\label{subsec:exp:diversity}
\begin{figure}[htp]
     \vspace{-0.18in}
        \begin{center}
    \begin{adjustwidth*}{-0.5em}{0em}
        \input{Images/adders}
    \end{adjustwidth*}
    \end{center}
      \vspace{-0.1in}
      \caption{Design diversity comparison for diverse 64-bit adder implementations synthesized with low and high area efforts versus average rareness for 100 most rare signals ($\mu(\omega_{100})$)} %Note that the higher $\mu(\omega_{100})$ value is better for Trojan Detection.}
        \vspace{-0.1in}
      \label{fig:adders}
\end{figure}
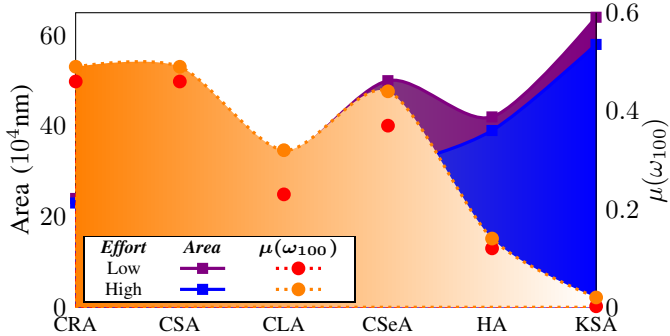

For this experiment, we have selected 64-bit adder circuits of CarryRipple (CRA), CarrySkip (CSA), CarryLookAhead (CLA), CarrySelect (CSeA), Hybrid (HA) and Kogge-Stone (KSA). These circuits were synthesized in two area effort levels of high and low. Next, we simulated the synthesized circuits individually with 10,000 randomly generated test patterns. Then using the VCD dump, we calculated the average rareness of the 100 most rare signals in each circuit. Figure~\ref{fig:adders} shows the results of the experiment. 

It can be observed that the average rareness for different algorithms that implement the same functionality are different. We analyzed the reason behind the drastically low $\mu(\omega_{100})$ values of the CarryLookAhead adder. In the synthesized design, it was observed that there are paths consisting of only NAND gates. This phenomenon negatively contributes to the rareness significantly, due to the involvement of the same type of gates in the propagation path of the signal, as discussed in Section~\ref{subsubsec:logicDepth}. Therefore, it is important to consider diversity of algorithms and their rareness metrics to build a design-for-trust solution. Although the average rareness varies across the algorithms due to design diversity, the relationship between the area and the rareness metrics still holds for each algorithm. This can be observed by observing the rareness metrics at different area synthesis efforts. The designs with the lowest area have the highest average rareness, while the designs with the higher area have the rarest signals. This means that the signals in the area-optimized design are less rare, making it easier for Trojan detection. 
%Further discussion about the relationship between area and rareness metrics is continued in Section~\ref{subsec:exp:area}. 

\begin{figure*}[htp]
\vspace{-0.1in}
\centering
\input{Images/correlation}
\vspace{-0.05in}
\caption{Correlation analysis heat-map generated by analyzing synthesized design features vs rareness metrics ($\mu(\omega_{all})$,$\rho_{(< 0.1)}$)}% and number of rare signals below the threshold of 0.1 ($\rho_{(< 0.1)}$) on selected set of real-world benchmark IPs}
        % \vspace{-0.15in}
      \label{fig:correlation}
\end{figure*}

\subsection{Area Optimization versus Rareness Correlation}\label{subsec:exp:area}
In order to empirically prove the hypothesis that we have outlined in Section~\ref{sec:methodology}, we have created a correlation analysis experiment. For this we have selected diverse designs covering network-on-chip (NoC) routers, processors (Attiny), crypto cores (AES and ECDSA), error correcting (ECC) memory cores from \textit{OpenCores}~\cite{opencores}. Figure~\ref{fig:correlation} presents the correlation heat-map for the design physical features against the average rareness ($\mu(\omega)$) and number of rare nodes below the rareness threshold of 0.1 ($\rho_{(<0.1)}$). The results were obtained using the experimental setup illustrated in Figure~\ref{fig:exp1}. First, we have synthesized the designs with three different area effort levels of low, medium and high. Then we have simulated the synthesized designs with 10,000 test patterns to calculate the rareness metrics. Finally, we have calculated the correlation coefficient value for each design parameter against the signal rareness metrics. It can be observed that the design area is  positively ($\color{cardinal}\blacksquare$) correlated with $\rho_{(<0.1)}$ ($A\propto \rho_{(<0.1)}$) while design area is negatively ($\color{deeplilac}\blacksquare$) correlated  with $\mu(\omega)$ ($A\propto \frac{1}{\mu(\omega)}$). This confirms that fact that the theoretical properties holds true on real-world designs. Further, it can be observed that the correlation between the rareness metrics against the logic levels varies depending on the design. This reflects the effect of gate type involved in the design for the signal rareness.

\begin{table}[h]
\centering
% \vspace{-0.1in}
\caption{Percentage comparison of area reduction ({A\bm{$\downarrow$}}) , effect on rareness metrics ($\rho_{(<0.1)}\downarrow$, $\Delta\mu(\omega_{all})\uparrow$) and test generation time reduction for different hardware designs. The complexity of the designs in terms of number of logic gates is as follows: ECC memory (100K), Attiny processor (30K), NoC router (10K), AES (80K), and ECDSA (300K).}
\vspace{-0.05in}
\label{tab:redPercent}
\begin{adjustbox}{max width=0.49\textwidth}
\begin{tabular}{|c|c|c|c|cc|}
\hline
\multicolumn{1}{|c|}{\multirow{2}{*}{\textbf{Design}}} & \multicolumn{1}{c|}{\multirow{2}{*}{A\bm{$\downarrow$}\%}} & \multicolumn{1}{c|}{\multirow{2}{*}{\bm{$\rho\downarrow$}\%}} & %_{(<0.1)}\downarrow$}\%}} & \footnotesize{$\Delta$}\bm{$\mu(\omega)\uparrow$}

\multicolumn{1}{c|}{\multirow{2}{*}{$\Delta\mu\uparrow$}} & \multicolumn{2}{c|}{Test Generation Time $\downarrow$} \\ \cline{5-6} 
\multicolumn{1}{|c|}{} & \multicolumn{1}{c|}{} & \multicolumn{1}{c|}{} & \multicolumn{1}{c|}{} & \multicolumn{1}{c|}{MERO\cite{mero2009swarup}} & \footnotesize{TARMAC\cite{tarmac2021yangdi}}\\ \hline
ECC mem & 10.1\% & 5.8\% & 0.007 & \multicolumn{1}{c|}{8.9\%} & 23\% \\ \hline
Attiny & 4.8\% & 3.4\% & 0.012 & \multicolumn{1}{c|}{7.2\%} & 19.8\% \\ \hline
NoC router & 7.3\% & 6.1\% & 0.010 & \multicolumn{1}{c|}{10.3\%} & 24.1\% \\ \hline
AES & 5.2\% & 11.8\% & 0.009 & \multicolumn{1}{c|}{5.8\%} & 17.9\% \\ \hline
ECDSA & 12.1\% & 9.7\% & 0.018 & \multicolumn{1}{c|}{13.6\%} & 28.4\% \\ \hline
\end{tabular}
\end{adjustbox}
\vspace{-0.1in}
\end{table}

Table~\ref{tab:redPercent} presents the percentage reduction of area between lowest and highest area effort setting with the decrement of $\rho$ and increment of $\omega$ for different benchmarks in this experiment.

\subsection{Effectiveness of Rareness Reduction on Trojan Detection}\label{subsec:exp:testgeneration}

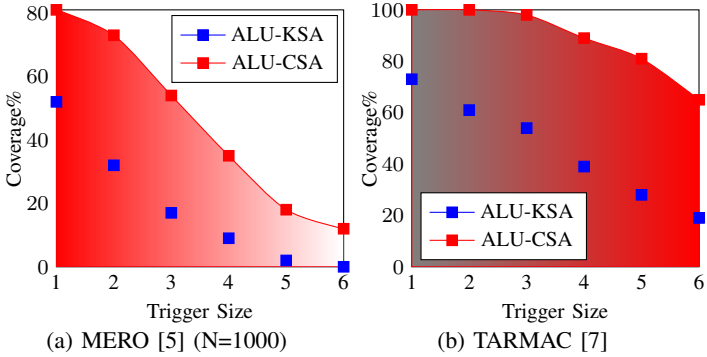
\begin{figure}[htp]
     \vspace{-0.18in}
        % \begin{center}
    \begin{adjustwidth*}{-2em}{-2em}
        \begin{subfigure}{.5\linewidth}
        %   \centering
          \input{Images/adderCovMero}
          \vspace{-0.25in}
          \caption{MERO~\cite{mero2009swarup} (N=1000)}
          \label{fig:diversity-meroCov}
        \end{subfigure}%
        \hspace{-1.5em}
        \begin{subfigure}{.5\linewidth}
        %   \centering
          \input{Images/adderCovTarmac}
          \vspace{-0.25in}
          \caption{TARMAC~\cite{tarmac2021yangdi}}
          \label{fig:diversity-tarmacCov}
        \end{subfigure}
    \end{adjustwidth*}
    % \end{center}
      \vspace{-0.05in}
      \caption{Design diversity coverage improvement for ALU-CSA and ALU-KSA designs with MERO and TARMAC.($\tau$ = 0.2)}
        % \vspace{-0.1in}($\tau$ = 0.2)
      \label{fig:coverage-adders}
\end{figure}

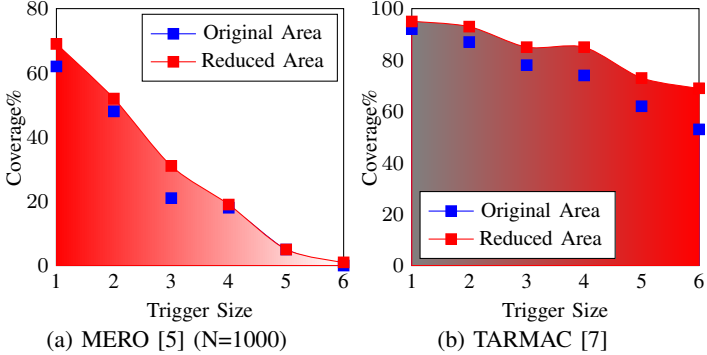
\begin{figure}[htp]
     \vspace{-0.18in}
        % \begin{center}
    \begin{adjustwidth*}{-2em}{-2em}

        \begin{subfigure}{.5\linewidth}
          \centering
          \input{Images/meroCoverage}
          \vspace{-0.25in}
          \caption{MERO~\cite{mero2009swarup} (N=1000)}
          \label{fig:area-meroCov}
        \end{subfigure}%
        \hspace{-1.5em}
        \begin{subfigure}{.5\linewidth}
          \centering
          \input{Images/tarmacCoverage}
          \vspace{-0.25in}
          \caption{TARMAC~\cite{tarmac2021yangdi}}
          \label{fig:area-tarmacCov}
        \end{subfigure}
    \end{adjustwidth*}
    % \end{center}
      \vspace{-0.05in}
    %   \caption{Yielded coverage improvement by MERO and TARMAC on ECDSA core module before and after rareness reduction for different trigger sizes.($\tau$ = 0.2)}
    \caption{Area optimization coverage improvement for ECDSA module with MERO and TARMAC.($\tau$ = 0.2)} %
        \vspace{-0.2in}
      \label{fig:coverage}
\end{figure}

For this experiment, we have used the MERO~\cite{mero2009swarup} and TARMAC~\cite{tarmac2021yangdi} test generation-based Trojan detection algorithms. First, we have generated test vectors ($\tau$= 0.2) using both methods on the design before and after rareness reduction. The test generation time reduction column of Table~\ref{tab:redPercent} illustrates the time saved during the test generation process by each method.  For MERO, we have used $N$ as 1000.  Then we randomly insert Trojans into the design following the method outlined in~\cite{cruz2018automated} to evaluate the coverage improvement. Specifically, we compute the \textit{Trojan coverage} as the ratio between the number of Trojans detected by the test vectors and the total number of inserted Trojans.

In order to evaluate the effectiveness of rareness reduction due to design diversity, we have used two 64-bit ALU implementations. For the first ALU, we have inserted an instance of CarrySkip adder (ALU-CSA) and for the second ALU, we replaced it with a Kogge-Stone adder (ALU-KSA). Figure~\ref{fig:coverage-adders} illustrates the coverage results of the design diversity experiment. It can be observed that although the functionality of the two ALU's still the same, different implementations yields drastically different Trojan coverage results. In this experiment, the ALU-CSA implementation is more friendly toward security verification.

In order to demonstrate the effectiveness of area optimization based rareness reduction, we have selected the ECDSA core as the evaluation benchmark. Figure~\ref{fig:coverage} demonstrates the coverage improvement results on the ECDSA benchmark before and after area optimization.  It can be observed that coverage has been improved in both MERO and TARMAC methods on the most area optimized design.

%% file: Images/empExperiments.tex
        \tikzset{every picture/.style={line width=0.75pt}} %set default line width to 0.75pt        

        \begin{tikzpicture}[x=0.75pt,y=0.75pt,yscale=-0.6,xscale=0.775]
        %uncomment if require: \path (0,0); %set diagram left start at 0, and has height of 419
        
        %Shape: Rectangle [id:dp6952430975237776] 
        \draw  [draw opacity=0][fill={rgb, 255:red, 153; green, 0; blue, 77 }  ,fill opacity=1 ][line width=0.75]  (45.34,33.5) -- (173.36,33.5) -- (173.36,80.5) -- (45.34,80.5) -- cycle ;
        
        %Shape: Rectangle [id:dp881435087966637] 
        \draw  [draw opacity=0][fill={rgb, 255:red, 153; green, 0; blue, 77 }  ,fill opacity=1 ][line width=0.75]  (44.98,163.77) -- (173,163.77) -- (173,210.77) -- (44.98,210.77) -- cycle ;
        
        %Shape: Rectangle [id:dp13826936249350852] 
        \draw  [draw opacity=0][fill={rgb, 255:red, 153; green, 0; blue, 77 }  ,fill opacity=1 ][line width=0.75]  (197.07,164.1) -- (325.08,164.1) -- (325.08,211.1) -- (197.07,211.1) -- cycle ;
        
        %Shape: Rectangle [id:dp46870860810889914] 
        \draw  [draw opacity=0][fill={rgb, 255:red, 153; green, 0; blue, 77 }  ,fill opacity=1 ][line width=0.75]  (351.75,33.43) -- (479.77,33.43) -- (479.77,80.43) -- (351.75,80.43) -- cycle ;
        
        %Shape: Rectangle [id:dp9097010663485201] 
        \draw  [draw opacity=0][fill={rgb, 255:red, 153; green, 0; blue, 77 }  ,fill opacity=1 ][line width=0.75]  (350.85,164.1) -- (478.87,164.1) -- (478.87,211.1) -- (350.85,211.1) -- cycle ;
        
        %Shape: Ellipse [id:dp24343444048668883] 
        \draw  [draw opacity=0][fill={rgb, 255:red, 120; green, 89; blue, 163 }  ,fill opacity=1 ][line width=0.75]  (355.09,120.5) .. controls (355.09,105.71) and (382,93.72) .. (415.2,93.72) .. controls (448.39,93.72) and (475.3,105.71) .. (475.3,120.5) .. controls (475.3,135.3) and (448.39,147.29) .. (415.2,147.29) .. controls (382,147.29) and (355.09,135.3) .. (355.09,120.5) -- cycle ;
        
        %Shape: Rectangle [id:dp677151317452948] 
        \draw  [draw opacity=0][fill={rgb, 255:red, 120; green, 89; blue, 163 }  ,fill opacity=1 ][line width=0.75]  (197.07,33.05) -- (325.08,33.05) -- (325.08,80.05) -- (197.07,80.05) -- cycle ;
        
        %Shape: Rectangle [id:dp5201604708933132] 
        \draw  [draw opacity=0][fill={rgb, 255:red, 120; green, 89; blue, 163 }  ,fill opacity=1 ][line width=0.75]  (197.07,97.1) -- (325.08,97.1) -- (325.08,144.1) -- (197.07,144.1) -- cycle ;
        
        %Straight Lines [id:da6302539496095498] 
        \draw [line width=0.75]    (110.12,101.7) -- (194.35,101.58) ;
        \draw [shift={(197.35,101.57)}, rotate = 179.92] [fill={rgb, 255:red, 0; green, 0; blue, 0 }  ][line width=0.08]  [draw opacity=0] (8.93,-4.29) -- (0,0) -- (8.93,4.29) -- cycle    ;
        %Straight Lines [id:da9023529957848875] 
        \draw [line width=0.75]    (110.07,120.75) -- (194.35,120.9) ;
        \draw [shift={(197.35,120.91)}, rotate = 180.11] [fill={rgb, 255:red, 0; green, 0; blue, 0 }  ][line width=0.08]  [draw opacity=0] (8.93,-4.29) -- (0,0) -- (8.93,4.29) -- cycle    ;
        %Straight Lines [id:da46080432384306325] 
        \draw [line width=0.75]    (184.8,138.06) -- (194.17,137.89) ;
        \draw [shift={(197.17,137.83)}, rotate = 178.97] [fill={rgb, 255:red, 0; green, 0; blue, 0 }  ][line width=0.08]  [draw opacity=0] (8.93,-4.29) -- (0,0) -- (8.93,4.29) -- cycle    ;
        %Straight Lines [id:da5888587205753915] 
        \draw [line width=0.75]    (184.8,138.06) -- (184.71,183.9) ;
        %Straight Lines [id:da4022207935377282] 
        \draw [line width=0.75]    (196.91,183.83) -- (184.71,183.9) ;
        %Straight Lines [id:da0027322403780658533] 
        \draw [line width=0.75]    (337.86,184.06) -- (328.45,184.39) ;
        \draw [shift={(325.45,184.5)}, rotate = 357.95] [fill={rgb, 255:red, 0; green, 0; blue, 0 }  ][line width=0.08]  [draw opacity=0] (8.93,-4.29) -- (0,0) -- (8.93,4.29) -- cycle    ;
        %Straight Lines [id:da7482619622232498] 
        \draw [line width=0.75]    (337.95,138.06) -- (337.86,184.06) ;
        %Straight Lines [id:da549467706857528] 
        \draw [line width=0.75]    (337.95,138.06) -- (325.19,138.17) ;
        %Straight Lines [id:da22795526037932734] 
        \draw [line width=0.75]    (337.91,74.2) -- (348.04,73.92) ;
        \draw [shift={(351.04,73.83)}, rotate = 178.41] [fill={rgb, 255:red, 0; green, 0; blue, 0 }  ][line width=0.08]  [draw opacity=0] (8.93,-4.29) -- (0,0) -- (8.93,4.29) -- cycle    ;
        %Straight Lines [id:da8146553226385576] 
        \draw [line width=0.75]    (173.32,56.64) -- (194.36,56.74) ;
        \draw [shift={(197.36,56.75)}, rotate = 180.25] [fill={rgb, 255:red, 0; green, 0; blue, 0 }  ][line width=0.08]  [draw opacity=0] (8.93,-4.29) -- (0,0) -- (8.93,4.29) -- cycle    ;
        %Straight Lines [id:da42316075904090145] 
        \draw [line width=0.75]    (337.91,74.2) -- (337.85,101.57) -- (324.65,101.57) ;
        %Straight Lines [id:da3816012547077956] 
        \draw [line width=0.75]    (415.99,80.44) -- (416.03,90.73) ;
        \draw [shift={(416.04,93.73)}, rotate = 269.78] [fill={rgb, 255:red, 0; green, 0; blue, 0 }  ][line width=0.08]  [draw opacity=0] (8.93,-4.29) -- (0,0) -- (8.93,4.29) -- cycle    ;
        %Straight Lines [id:da8581940220105528] 
        \draw [line width=0.75]    (110.05,80.39) -- (110.12,101.7) ;
        %Straight Lines [id:da38856655670660656] 
        \draw [line width=0.75]    (325.11,56.49) -- (348.71,56.67) ;
        \draw [shift={(351.71,56.7)}, rotate = 180.45] [fill={rgb, 255:red, 0; green, 0; blue, 0 }  ][line width=0.08]  [draw opacity=0] (8.93,-4.29) -- (0,0) -- (8.93,4.29) -- cycle    ;
        %Straight Lines [id:da6491605359633863] 
        \draw [line width=0.75]    (415.69,147.31) -- (415.78,160.2) ;
        \draw [shift={(415.8,163.2)}, rotate = 269.58] [fill={rgb, 255:red, 0; green, 0; blue, 0 }  ][line width=0.08]  [draw opacity=0] (8.93,-4.29) -- (0,0) -- (8.93,4.29) -- cycle    ;
        %Straight Lines [id:da6244257321213988] 
        \draw [line width=0.75]    (110.07,120.75) -- (109.95,163.91) ;
        
        % Text Node
        \draw (109.35,57.15) node   [align=left] {\begin{minipage}[lt]{87.05pt}\setlength\topsep{0pt}
        \begin{center}
        \textcolor[rgb]{1,1,1}{\textbf{Design}}\\\textcolor[rgb]{1,1,1}{{\footnotesize (RTL/GateLevel)}}
        \end{center}
        
        \end{minipage}};
        % Text Node
        \draw (108.99,187.22) node   [align=left] {\begin{minipage}[lt]{87.05pt}\setlength\topsep{0pt}
        \begin{center}
        \textcolor[rgb]{1,1,1}{\textbf{Technology}}\\\textcolor[rgb]{1,1,1}{{\footnotesize (saed90nm)}}
        \end{center}
        
        \end{minipage}};
        % Text Node
        \draw (261.08,187.55) node   [align=left] {\begin{minipage}[lt]{87.05pt}\setlength\topsep{0pt}
        \begin{center}
        \textcolor[rgb]{1,1,1}{\textbf{Synth. Script}}\\\textcolor[rgb]{1,1,1}{{\scriptsize (effort High/Med/Low)}}
        \end{center}
        
        \end{minipage}};
        % Text Node
        \draw (415.76,56.88) node   [align=left] {\begin{minipage}[lt]{87.05pt}\setlength\topsep{0pt}
        \begin{center}
        \textcolor[rgb]{1,1,1}{\textbf{Netlist}}\\\textcolor[rgb]{1,1,1}{{\footnotesize (GateLevel)}}
        \end{center}
        
        \end{minipage}};
        % Text Node
        \draw (414.86,187.55) node   [align=left] {\begin{minipage}[lt]{87.05pt}\setlength\topsep{0pt}
        \begin{center}
        \textcolor[rgb]{1,1,1}{\textbf{Metrics}}\\\textcolor[rgb]{1,1,1}{{\footnotesize $\displaystyle ( \mu ( \omega ) ,\rho ,\Omega )$}}
        \end{center}
        
        \end{minipage}};
        % Text Node
        \draw (415.2,120.5) node   [align=left] {\begin{minipage}[lt]{53.08pt}\setlength\topsep{0pt}
        \begin{center}
        \textcolor[rgb]{1,1,1}{\textbf{Simulation}}
        \end{center}
        
        \end{minipage}};
        % Text Node
        \draw (261.08,56.5) node   [align=left] {\begin{minipage}[lt]{87.05pt}\setlength\topsep{0pt}
        \begin{center}
        \textcolor[rgb]{1,1,1}{\textbf{Test Vectors}}\\\textcolor[rgb]{1,1,1}{{\footnotesize (Random)}}
        \end{center}
        
        \end{minipage}};
        % Text Node
        \draw (261.08,120.55) node   [align=left] {\begin{minipage}[lt]{87.05pt}\setlength\topsep{0pt}
        \begin{center}
        \textcolor[rgb]{1,1,1}{\textbf{Compiler}}\\\textcolor[rgb]{1,1,1}{{\footnotesize (DC\_shell)}}
        \end{center}
        
        \end{minipage}};

        \end{tikzpicture}

%% file: Images/adders.tex
\begin{tikzpicture}

\begin{axis}[
        %title = {},
        % xlabel= X LABEL HERE, 
        ylabel= {Area ($10^{4}$nm)},
        ymin=0, ymax=65,
        xtick style={draw=none},
        ytick style={draw=none},
        xtick=data,
        width=8.5cm,
        height=5.5cm,
        xmin=CRA,
        xmax=KSA,
        symbolic x coords={CRA,CSA,CLA,CSeA,HA,KSA},
        x tick label style={font=\footnotesize,text width=2cm,align=center},
        xticklabel style={rotate=0},
        y label style={at={(axis description cs:-.15,0.5)},anchor=north},
]

\addplot+[smooth,mark=square*,mark options={scale=0.75, fill=violet,fill opacity=1}  ,violet, right color=violet,fill opacity=0.1, draw opacity=1,very thick]
  coordinates{
    (CRA,24)
    (CSA,27)
    (CLA,32)
    (CSeA,50)
    (HA,42)
    (KSA,64)
}\closedcycle; \label{A}

\addplot+[smooth,mark=square*,mark options={scale=0.75, fill=blue,fill opacity=1},blue, right color=blue,fill opacity=0.1, draw opacity=1,very thick]
  coordinates{
    (CRA,23)
    (CSA,22)
    (CLA,29)
    (CSeA,31)
    (HA,39)
    (KSA,58)
}\closedcycle; \label{B}

\end{axis}

\begin{axis}[
  axis y line*=right,
  axis x line=none,
  ymin=0, ymax=0.6,
    ylabel= {$\mu(\omega_{100})$},
    % legend style={
    %         at={(0.5,1.2)},               
    %         anchor=north,legend columns=-2,
    %         draw=none
    % },
    ytick style={draw=none},
    xtick style={draw=none},
    xtick=data,
    width=8.5cm,
    height=5.5cm,
    xmin=CRA,
    xmax=KSA,
    symbolic x coords={CRA,CSA,CLA,CSeA,HA,KSA},
    x tick label style={font=\footnotesize,text width=2cm,align=center},
    y label style={at={(axis description cs:1.07,0.5)},anchor=north},
]
% \addlegendimage{/pgfplots/refstyle=A}\addlegendentry{Effort Low Area}
% \addlegendimage{/pgfplots/refstyle=B}\addlegendentry{plot 2}
% \addlegendimage{/pgfplots/refstyle=C}\addlegendentry{plot 3}
% \addlegendimage{/pgfplots/refstyle=D}\addlegendentry{plot 4}

\addplot+[smooth,mark=*,mark options={scale=1, fill=red,fill opacity=1,solid},red, left color=red,fill opacity=0.1, draw opacity=1,very thick,dotted]
coordinates{
    (CRA,0.46)
    (CSA,0.46)
    (CLA,0.23)
    (CSeA,0.37)
    (HA,0.12)
    (KSA,0.002)
}\closedcycle; \label{C}

\addplot+[smooth,mark=*,mark options={scale=1, fill=orange,fill opacity=1,solid},orange, left color=orange,fill opacity=0.1, draw opacity=1,very thick,dotted]
  coordinates{
    (CRA,0.49)
    (CSA,0.49)
    (CLA,0.32)
    (CSeA,0.44)
    (HA,0.14)
    (KSA,0.02)
}\closedcycle; \label{D}

% \legend{Area(Low effort), Area(High effort), C, D}

\end{axis}

% \draw (3.2,4.75) node   [align=center]
% \draw (3.2,4.92) node   [align=center]
\draw (3.85,0.5) node   [align=center]
{\begin{minipage}[lt]{212.84pt}\setlength\topsep{0pt}
% \begin{center}
% \footnotesize{\textbf{Low Effort :} \ref{A} Area \ref{C} $\mu(\omega_{100})$\\
% \textbf{High Effort :} \ref{B} Area \ref{D} $\mu(\omega_{100})$}
\scriptsize
\begin{tabular}{|ccc|}
\hline
\cellcolor{white}\textbf{\textit{Effort}} & \cellcolor{white}\textbf{\textit{Area}} & \cellcolor{white}\textit{\bm{$\mu(\omega_{100})$}} \\
\cellcolor{white}Low & \cellcolor{white}\ref{A} & \cellcolor{white}\ref{C} \\
\cellcolor{white}High & \cellcolor{white}\ref{B} & \cellcolor{white}\ref{D} \\ \hline
\end{tabular}
% \end{center}

\end{minipage}};

\end{tikzpicture}

%% file: Images/correlation.tex
\begin{adjustbox}{max width=\textwidth}
%\footnotesize
\small
\begin{tabular}{cccccccccc}
 & Metrics & Logic Levels & Leaf Cells & Comb.Cells & Comb.Area & Net Area & Cell Area & Design Area & Total Nets \\ \hline
\multirow{2}{*}{\begin{tabular}[c]{@{}c@{}}ECC\\ memory\end{tabular}} & $\mu(\omega_{all})$ & 
\color{black!69.3360171001217}\cellcolor{cardinal!21.902844928484505}{0.44}& 
 \color{white}\cellcolor{deeplilac!51.34880562408702}{-0.68}& 
 \color{white}\cellcolor{deeplilac!51.34880562408702}{-0.68}& 
 \color{white}\cellcolor{deeplilac!63.44907986164816}{-0.85}& 
 \color{white}\cellcolor{deeplilac!62.37670466847224}{-0.83}& 
 \color{white}\cellcolor{deeplilac!63.45510161298169}{-0.85}& 
 \color{white}\cellcolor{deeplilac!67.7640737387803}{-0.90}& 
 \color{white}\cellcolor{deeplilac!48.26269632063419}{-0.64} \\
 & $\rho_{(< 0.1)}$ & 
 \color{white}\cellcolor{deeplilac!55.3233268502051}{-0.74}& 
\color{white}\cellcolor{cardinal!67.83933500643023}{0.90}& 
\color{white}\cellcolor{cardinal!67.83933500643023}{0.90}& 
\color{white}\cellcolor{cardinal!73.70301888291004}{0.98}& 
\color{white}\cellcolor{cardinal!42.71910604061998}{0.57}& 
\color{white}\cellcolor{cardinal!73.70510950006404}{0.98}& 
\color{white}\cellcolor{cardinal!74.83301990989638}{1.00}& 
\color{white}\cellcolor{cardinal!65.97654778245544}{0.88} \\ \hline

\multirow{2}{*}{\begin{tabular}[c]{@{}c@{}}Attiny\\ Core\end{tabular}} & $\mu(\omega_{all})$ &
\color{white}\cellcolor{cardinal!60.18517392726757}{0.80}& 
\color{black!69.30744480841854}\cellcolor{deeplilac!21.92325370827247}{-0.44}& 
 \color{white}\cellcolor{deeplilac!42.34880562408702}{-0.56}& 
 \color{white}\cellcolor{deeplilac!51.44907986164816}{-0.69}& 
\color{black!78.1817423094259}\cellcolor{deeplilac!15.5844697789815}{-0.31}& 
 \color{white}\cellcolor{deeplilac!65.7051016129817}{-0.88}& 
 \color{white}\cellcolor{deeplilac!65.5140737387803}{-0.87}& 
 \color{white}\cellcolor{deeplilac!66.26269632063419}{-0.88} \\
 & $\rho_{(< 0.1)}$ & 
\color{black!79.43648949398086}\cellcolor{deeplilac!14.688221790013673}{-0.29}& 
\color{white}\cellcolor{cardinal!59.28393350064302}{0.79}& 
\color{white}\cellcolor{cardinal!58.53393350064302}{0.78}& 
\color{white}\cellcolor{cardinal!61.70301888291004}{0.82}& 
\color{white}\cellcolor{cardinal!61.46910604061998}{0.82}& 
\color{white}\cellcolor{cardinal!68.45510950006404}{0.91}& 
\color{white}\cellcolor{cardinal!68.83301990989638}{0.92}& 
\color{white}\cellcolor{cardinal!59.97654778245544}{0.80} \\ \hline

\multirow{2}{*}{\begin{tabular}[c]{@{}c@{}}ProNoC\\ router\end{tabular}} & $\mu(\omega_{all})$ &
\color{black!99.82717100121693}\cellcolor{cardinal!0.12344928484505}{0.00}& 
\color{black!83.30744480841855}\cellcolor{deeplilac!11.923253708272467}{-0.24}& 
 \color{white}\cellcolor{deeplilac!63.4880562408702}{-0.85}& 
 \color{white}\cellcolor{deeplilac!54.8616481575}{-0.73}& 
\color{white}\cellcolor{cardinal!41.847225}{0.56}& 
 \color{white}\cellcolor{deeplilac!45.5101612981695}{-0.61}& 
 \color{white}\cellcolor{deeplilac!73.7387802975}{-0.98}& 
 \color{white}\cellcolor{deeplilac!63.4185}{-0.85}\\
 & $\rho_{(< 0.1)}$ & 
\color{black!86.43648949398086}\cellcolor{deeplilac!9.688221790013674}{-0.19}& 
\color{black!93.66832873273317}\cellcolor{cardinal!4.5226223337620155}{0.09}& 
\color{white}\cellcolor{cardinal!69.03393350064302}{0.92}& 
\color{white}\cellcolor{cardinal!51.88829100375}{0.69}& 
 \color{white}\cellcolor{deeplilac!60.604061997749994}{-0.81}& 
\color{white}\cellcolor{cardinal!55.109500064040006}{0.73}& 
\color{white}\cellcolor{cardinal!65.989638}{0.88}& 
\color{white}\cellcolor{cardinal!57.4554375}{0.77} \\ \hline

\multirow{2}{*}{\begin{tabular}[c]{@{}c@{}}AES\\ Core\end{tabular}} & $\mu(\omega_{all})$ &
\color{white}\cellcolor{cardinal!42.673927267574996}{0.57}& 
\color{black!88.4185448}\cellcolor{deeplilac!8.272468}{-0.17}& 
 \color{white}\cellcolor{deeplilac!72.380562408702}{-0.97}& 
 \color{white}\cellcolor{deeplilac!74.0798616481575}{-0.99}& 
\color{black!88.17423094259}\cellcolor{deeplilac!8.446977898150001}{-0.17}& 
 \color{white}\cellcolor{deeplilac!60.1612981695}{-0.80}& 
 \color{white}\cellcolor{deeplilac!62.38780297500001}{-0.83}& 
 \color{white}\cellcolor{deeplilac!63.4185}{-0.85} \\
 & $\rho_{(< 0.1)}$ & 
\color{black!80.8578}\cellcolor{deeplilac!13.672999999999998}{-0.27}& 
 \color{white}\cellcolor{deeplilac!39.302325}{-0.52}& 
\color{white}\cellcolor{cardinal!70.89335006430233}{0.95}& 
\color{white}\cellcolor{cardinal!63.29100375}{0.84}& 
\color{white}\cellcolor{cardinal!40.619977500000005}{0.54}& 
\color{white}\cellcolor{cardinal!73.70510950006404}{0.98}& 
\color{white}\cellcolor{cardinal!65.989638}{0.88}& 
\color{white}\cellcolor{cardinal!65.477824554375}{0.87} \\ \hline

\multirow{2}{*}{\begin{tabular}[c]{@{}c@{}}ECDSA\\ Sign\end{tabular}} & $\mu(\omega_{all})$ &
\color{black!74.17100121693}\cellcolor{deeplilac!18.449284845050002}{-0.37}& 
\color{black!70.84185448}\cellcolor{cardinal!20.827246799999998}{0.42}& 
 \color{white}\cellcolor{deeplilac!55.624087020000005}{-0.74}& 
 \color{white}\cellcolor{deeplilac!74.0798616481575}{-0.99}& 
 \color{white}\cellcolor{deeplilac!51.70466847225}{-0.69}& 
 \color{white}\cellcolor{deeplilac!66.01612981695}{-0.88}& 
 \color{white}\cellcolor{deeplilac!67.7387802975}{-0.90}& 
 \color{white}\cellcolor{deeplilac!73.12696320634186}{-0.98} \\
 & $\rho_{(< 0.1)}$ & 
\color{black!79.43648949398086}\cellcolor{cardinal!14.688221790013673}{0.29}& 
\color{black!82.87327331783}\cellcolor{cardinal!12.23337620155}{0.24}& 
\color{white}\cellcolor{cardinal!50.64302325}{0.68}& 
\color{white}\cellcolor{cardinal!68.8829100375}{0.92}& 
\color{white}\cellcolor{cardinal!60.406199775}{0.81}& 
\color{white}\cellcolor{cardinal!59.500064040000005}{0.79}& 
\color{white}\cellcolor{cardinal!67.3301990989638}{0.90}& 
\color{white}\cellcolor{cardinal!68.97654778245544}{0.92} \\ \hline

\end{tabular}

\end{adjustbox}

%% file: Images/adderCovMero.tex
\begin{tikzpicture}
\footnotesize
\begin{axis}[
        xlabel= Trigger Size, 
        ylabel= {Coverage\%},
        ymin=0, ymax=81,
        xtick style={draw=none},
        ytick style={draw=none},
        xtick=data,
        width=5.4cm,
        height=5cm,
        xmin=1,
        xmax=6,
        symbolic x coords={1,2,3,4,5,6},
        x tick label style={font=\footnotesize,text width=2cm,align=center},
        y label style={at={(axis description cs:-.2,0.5)},anchor=north},
]

\addplot+[smooth,mark=square*,mark options={scale=1, fill=blue,fill opacity=1}  ,blue, right color=blue,fill opacity=0.1, draw opacity=1 ]
  coordinates{
    (1,52)
    (2,32)
    (3,17)
    (4,9)
    (5,2)
    (6,0)
}\closedcycle; \label{E}

\addplot+[smooth,mark=square*,mark options={scale=1, fill=red,fill opacity=1},red, left color=red,fill opacity=0.1, draw opacity=1 ]
  coordinates{
    (1,81)
    (2,73)
    (3,54)
    (4,35)
    (5,18)
    (6,12)
}\closedcycle; \label{F}
\addlegendimage{/pgfplots/refstyle=E}\addlegendentry{ALU-KSA}
\addlegendimage{/pgfplots/refstyle=F}\addlegendentry{ALU-CSA}
\end{axis}

\end{tikzpicture}

%% file: Images/adderCovTarmac.tex
\begin{tikzpicture}
\footnotesize
\begin{axis}[
        xlabel= Trigger Size, 
        ylabel= {Coverage\%},
        ymin=0, ymax=100,
        xtick style={draw=none},
        ytick style={draw=none},
        xtick=data,
        width=5.4cm,
        height=5cm,
        xmin=1,
        xmax=6,
        symbolic x coords={1,2,3,4,5,6},
        x tick label style={font=\footnotesize,text width=2cm,align=center},
        y label style={at={(axis description cs:-.2,0.5)},anchor=north},
        legend style={
            at={(0.61,0.29)},               
    },
]

\addplot+[smooth,mark=square*,mark options={scale=1, fill=blue,fill opacity=1}  ,blue, left color=blue,fill opacity=0.1, draw opacity=1]
  coordinates{
    (1,73)
    (2,61)
    (3,54)
    (4,39)
    (5,28)
    (6,19)
}\closedcycle; \label{G}

\addplot+[smooth,mark=square*,mark options={scale=1, fill=red,fill opacity=1},red, right color=red,fill opacity=0.1, draw opacity=1]
  coordinates{
    (1,100)
    (2,100)
    (3,98)
    (4,89)
    (5,81)
    (6,65)
}\closedcycle; \label{H}
\addlegendimage{/pgfplots/refstyle=G}\addlegendentry{ALU-KSA}
\addlegendimage{/pgfplots/refstyle=H}\addlegendentry{ALU-CSA}

\end{axis}

\end{tikzpicture}

%% file: Images/meroCoverage.tex
\begin{tikzpicture}
\footnotesize
\begin{axis}[
        xlabel= Trigger Size, 
        ylabel= {Coverage\%},
        ymin=0, ymax=80,
        xtick style={draw=none},
        ytick style={draw=none},
        xtick=data,
        width=5.4cm,
        height=5cm,
        xmin=1,
        xmax=6,
        symbolic x coords={1,2,3,4,5,6},
        x tick label style={font=\footnotesize,text width=2cm,align=center},
        y label style={at={(axis description cs:-.2,0.5)},anchor=north},
]

\addplot+[smooth,mark=square*,mark options={scale=1, fill=blue,fill opacity=1}  ,blue, right color=blue,fill opacity=0.1, draw opacity=1 ]
  coordinates{
    (1,62)
    (2,48)
    (3,21)
    (4,18)
    (5,5)
    (6,0)
}\closedcycle; \label{I}

\addplot+[smooth,mark=square*,mark options={scale=1, fill=red,fill opacity=1},red, left color=red,fill opacity=0.1, draw opacity=1 ]
  coordinates{
    (1,69)
    (2,52)
    (3,31)
    (4,19)
    (5,5)
    (6,1)
}\closedcycle; \label{J}
\addlegendimage{/pgfplots/refstyle=I}\addlegendentry{Original Area}
\addlegendimage{/pgfplots/refstyle=J}\addlegendentry{Reduced Area}
\end{axis}

\end{tikzpicture}

%% file: Images/tarmacCoverage.tex
\begin{tikzpicture}
\footnotesize
\begin{axis}[
        xlabel= Trigger Size, 
        ylabel= {Coverage\%},
        ymin=0, ymax=100,
        xtick style={draw=none},
        ytick style={draw=none},
        xtick=data,
        width=5.4cm,
        height=5cm,
        xmin=1,
        xmax=6,
        symbolic x coords={1,2,3,4,5,6},
        x tick label style={font=\footnotesize,text width=2cm,align=center},
        y label style={at={(axis description cs:-.2,0.5)},anchor=north},
        legend style={
           at={(0.71,0.29)},               
    },
]

\addplot+[smooth,mark=square*,mark options={scale=1, fill=blue,fill opacity=1}  ,blue, left color=blue,fill opacity=0.1, draw opacity=1]
  coordinates{
    (1,92)
    (2,87)
    (3,78)
    (4,74)
    (5,62)
    (6,53)
}\closedcycle; \label{K}

\addplot+[smooth,mark=square*,mark options={scale=1, fill=red,fill opacity=1},red, right color=red,fill opacity=0.1, draw opacity=1]
  coordinates{
    (1,95)
    (2,93)
    (3,85)
    (4,85)
    (5,73)
    (6,69)
}\closedcycle; \label{L}
\addlegendimage{/pgfplots/refstyle=K}\addlegendentry{Original Area}
\addlegendimage{/pgfplots/refstyle=L}\addlegendentry{Reduced Area}

\end{axis}

\end{tikzpicture}

%% file: sections/conclusion.tex
\section{Conclusion} 
\label{sec:conclusion}

Design-for-trust is an important objective to develop secure and trustworthy systems. While obfuscation is a promising avenue, it can lead to unacceptable hardware overhead. In this paper, we explored the effectiveness of rareness reduction to design trustworthy systems. We performed a theoretical analysis of the root causes of rare signals that are likely to be exploited by adversaries to construct stealthy triggers in hardware Trojans. We also explored two techniques for rareness reduction, including design diversity, and area optimization. We performed empirical evaluation using real-world hardware benchmarks to demonstrate the validity of the theoretical analysis. We also conducted experiments to evaluate the effectiveness of rareness reduction for Trojan detection using statistical test generation as well as maximal clique activation. Experimental results demonstrated that our proposed rareness reduction techniques improved the Trojan detection efficiency in terms of reduction in test generation time as well as improved Trojan coverage. 
%Unlike the unacceptable area overheads added by logic locking-based approaches, the proposed technique leads to better Design-for-Trust that is benefited by state-of-the-art Trojan detection techniques and yields improved Trojan coverage.  